\documentclass[useAMS,usenatbib]{mn2e}

\usepackage{amsmath}
\usepackage{relsize}
\usepackage{graphicx}
\usepackage{epstopdf}
\usepackage{graphics}
\usepackage{multirow}
\usepackage{booktabs}
\usepackage{supertabular}
\usepackage{lscape}
\usepackage{longtable}
\usepackage[figuresright]{rotating}
\usepackage{multicol}
\usepackage{natbib}
\usepackage{rotating}
\usepackage{epsfig}
\usepackage{amsfonts}
\usepackage{threeparttable}
\usepackage[english]{babel}
\usepackage{textcomp}
\graphicspath{ {./graphics/} }
\usepackage{natbib}
\usepackage{subfigure}
\usepackage{tabularx}
\usepackage[width=1.0\textwidth]{caption}
\usepackage{pifont}
\usepackage{float}

\usepackage{url}  
\usepackage{amssymb}

\newcolumntype{R}[1]{>{\raggedleft\arraybackslash}p{#1}}
\newcolumntype{C}[1]{>{\centering\arraybackslash}p{#1}}
\newcolumntype{L}[1]{>{\raggedright\arraybackslash}p{#1}}
\newcommand{\kms}{${\rm km\,s}^{-1}$}
\newcommand{\VK}{$V-K_{\rm s}$}
\newcommand{\LxLbol}{$L_{\rm X}/L_{\rm bol}$}
\newcommand{\MK}{$M_{\rm K_{\rm s}}$}
\newcommand{\MV}{$M_{\rm V}$}

\title[The Pisces Moving Group]{Searching for new young stars in the northern hemisphere: The Pisces Moving Group}
\author[A. S. Binks, R. D. Jeffries and J. L. Ward]{A. S. Binks$^{1}$\thanks{E-mail: a.binks@crya.unam.mx}, R. D. Jeffries$^{2}$ and J. L. Ward$^{3}$\\
$^{1}$Instituto de Radioastronom\'ia y Astrof\'isica, Universidad Nacional Aut\'onoma de M\'exico, PO Box 3-72, 58090 Morelia, Michoac\'an, M\'exico\\
$^{2}$Astrophysics Group, School of Chemistry and Physics, Keele University, Keele, Staffordshire ST5 5BG\\
$^{3}$Astronomisches Rechen-Institut, Zentrum f\"{u}r Astronomie der Universit\"{a}t Heidelberg, M\"{o}nchhofstra{\ss}e 12-14, D-69120 Heidelberg, Germany}

\begin{document}

\date{Accepted. Received, in original form}

\pagerange{\pageref{firstpage}--\pageref{lastpage}} \pubyear{2017}

\maketitle

\label{firstpage}

\begin{abstract}
Using the kinematically unbiased technique described in Binks, Jeffries \& Maxted (2015), we present optical spectra for a further 122 rapidly-rotating (rotation periods $<6$~days), X-ray active FGK stars, selected from the SuperWASP survey. We identify 17 new examples of young, probably single stars with ages of $<200$\,Myr and provide additional evidence for a new northern hemisphere kinematic association: the Pisces Moving Group (MG). The group consists of 14 lithium-rich G- and K-type stars, that have a dispersion of only $\sim 3$ km\,s$^{-1}$ in each Galactic space velocity coordinate. The group members are approximately co-eval in the colour-magnitude diagram, with an age of 30--50\,Myr, and have similar, though not identical, kinematics to the Octans-Near MG.
\end{abstract}

\begin{keywords}
stars: low-mass -- stars: pre-main-sequence
\end{keywords}

\section{Introduction}\label{S_Intro}
Historically, the majority of young stars in the Solar neighbourhood were found to be part of a large kinematic structure known as the Local Association (LA). Discovered by Eggen in the 1960s (Eggen 1961; 1965a) by identifying samples of stars with approximately similar space velocities and ages, its constituents include large OB associations such as Scorpius Centaurus, open clusters such as IC~2602, $\alpha$~Persei and the Pleiades, and $\sim$1/3 of all B-type stars near the Sun (\citealt{1983a_Eggen}). A unitary view of the LA, however, was difficult to defend because its members spanned a large range of ages ($10-100$\,Myr) and are spatially distributed over $\sim 300\,$pc. Motivated by Eggen's hypothesis, detailed observations of nearby chromospherically active, rapidly-rotating, low-mass stars suggested that they were also part of the LA. Follow-up optical spectroscopy revealed a large fraction of these were indeed young and co-moving (\citealt{1988a_Innis, 1993a_Jeffries, 1995a_Jeffries}).

\nocite{1961a_Eggen}
\nocite{1965a_Eggen}
\nocite{1983a_Eggen}

Subsequent work suggested kinematic sub-structure within the LA in the form of several groups of co-moving, approximately co-eval, young stars known as nearby young moving groups (hereafter MGs, \citealt{2004a_Zuckerman, 2008a_Torres, 2013a_Malo}). Much recent work has focused on identifying and observing MGs and their members, as they provide crucial observational constraints for evolutionary stellar models and are excellent laboratories for testing the conditions and history of our nearest, young stars. A combination of youth, proximity and enhanced brightness contrasts makes MG members prime targets for directly imaging circumstellar discs and exoplanets (see \citealt{2013a_Biller, 2013a_Dent, 2014a_Brandt, 2015a_Bowler}). The ages of MG members ($\sim 5-150\,$Myr) are coincident with the epochs of pre-main sequence (PMS) evolution for solar-type and low-mass stars (\citealt{2015a_Bell}), the formation and dispersal of protoplanetary, transitional and debris discs (Wyatt et al. 2008; Boucher et al. 2016; Binks $\&$ Jeffries 2017) and the timescales on which giant and terrestrial planets are formed (\citealt{2008a_Morishima}).

\nocite{2008a_Wyatt}
\nocite{2016a_Binks}
\nocite{2016a_Boucher}

Techniques to identify new MGs and existing members of MGs can broadly be split into two philosophies. The first is to use the kinematics and ages of pre-existing MGs as preliminary search criteria for additional members (see, for example, \citealt{2006a_Torres, 2009a_Lepine, 2010a_Schlieder, 2012a_Schlieder, 2013a_Malo, 2014a_Elliott, 2017a_Riedel}). Follow-up observations of MG candidates are then made to further test membership status. These methods have proven effective at revealing, in particular, the low-mass counterparts of MGs (\citealt{2015a_Gagne}), however they can not identify new MGs and have an inherent bias towards limiting the location of any new MG members to the spatial extent of the currently known members. The second technique utilises large, often all-sky astronomical surveys which contain crude stellar youth-indicators such as activity, rotation, UV, X-ray or infrared excess and follow-up observations are made to measure their kinematics and ages (see da Silva et al. 2009; Kiss et al. 2011; Shkolnik et al. 2012; Binks, Jeffries $\&$ Maxted 2015 -- hereafter BJM15, Kastner et al. 2017). This technique is usually less efficient (generally $\lesssim 15$ per cent), however, it provides the opportunity to identify entirely new kinematically-coherent, young co-eval ensembles.

\nocite{2009a_da_Silva}
\nocite{2011a_Kiss}
\nocite{2012a_Shkolnik}
\nocite{2015a_Binks}
\nocite{2017a_Kastner}

The majority of searches for MGs and new members of existing MGs have been based in the southern hemisphere, presumably because this is where the majority of the known members of MGs have been located. The northern hemisphere remains lightly studied and potentially harbours many undiscovered members of existing MGs or new MGs. Photometric monitoring of many FGK stars in MGs suggest their rotation periods are several days or less (Messina et al. 2010, 2016). Using this as a starting point to search for young stars, BJM15 conducted a spectroscopic follow-up of short period, X-ray active solar-type stars in the northern hemisphere identified in both the SuperWASP All-Sky-Survey and the ROSAT 1RXP and 2RXS catalogues. Among the 26 stars identified as young and lithium-rich, there was evidence for a new group of 7 G/K stars that shared a common age and space motion. This grouping of stars in velocity space was labelled the `Pisces MG'; similar in kinematics to the Octans and the Octans-Near association (\citealt{2008a_Torres, 2013a_Zuckerman}), but quite separate from the space velocity of the LA MGs.

Motivated by the fact that our first search may have revealed a new kinematic structure of young stars we revisited the SuperWASP catalogue to search for more potentially young stars. In this paper we present spectroscopic data for 122 new targets, 103 of which were chosen based on the methodology of BJM15 and 19 of which had positions and proper motions that could {\it potentially} be co-moving with the Pisces MG (see $\S$\ref{SS_Potential_Pisces}). The first release of astrometric data from the {\it Gaia} mission on September 14, 2016, which contains sub-milliarcsecond positions, proper-motions and parallaxes for $\sim\,2$ million $V \lesssim 12$ stars in the Tycho astrometric catalogue (known as the Tycho-Gaia Astrometric Solution, TGAS, \citealt{2016a_GAIA_Collaboration}), is already proving an excellent tool for astronomers studying young, nearby stars and we improve on our results from BJM15 by incorporating TGAS data where appropriate.

\nocite{2015a_Binks}
\nocite{2010a_Messina}
\nocite{2016a_Messina}

The aims of the present paper are:

\begin{enumerate}
\item To expand the sample of spectroscopically confirmed young stars in the northern hemisphere, some of which are relatively close to the Sun and provide excellent targets for further study.

\item To search for additional members of the tentative Pisces MG identified in BJM15 and to confirm its reality as a co-eval, comoving group of stars.

\end{enumerate}

A detailed description of the initial target selection, data reduction and subsequent analyses are provided in BJM15. In $\S$\ref{S_Target_Selection} we describe how we cross-matched the SuperWASP and ROSAT catalogues to generate our initial catalogue of short-period, X-ray active FGK stars. We provide a description of the spectroscopic observations and data reduction in $\S$\ref{S_Observations} and in $\S$\ref{S_Data} how we extract radial velocities and equivalent widths from our spectra. In $\S$\ref{S_Ages} we provide the rationale for our estimates of the ages of stars in our sample and present the kinematic data in $\S$\ref{S_Kinematics}. A discussion of the existence of the Pisces MG and/or any other potential new groups is given in $\S$\ref{S_Pisces}. We summarise and conclude in $\S$\ref{S_Conclusion}.

\section{Target selection}\label{S_Target_Selection}

Target selection is based on two separate lists, both of which originate from a parent sample based on a cross correlation of targets present in both the SuperWASP all-sky-survey archive and either the ROSAT sky-survey (1RXS) or pointed-phase (2RXP) catalogues (Voges et al. 1999, 2000; Pollacco et al. 2006; Norton et al. 2007; Boller et al. 2016 -- see BJM15 for a detailed description of the parent sample from which we choose our targets). The initial sample contains 777 spectroscopically unobserved SuperWASP/ROSAT objects, with rotation periods $\lesssim 5$ days, declinations $> -20^{\circ}$, $8 < V < 15$ and spectral-types FGK, corresponding to $0.7 < V-K_{\rm s} < 3.8$.

Targets in the first list are selected from the parent sample. A second list comprised of a subset of 160 targets whose proper-motions and positions are compatible with membership of the Pisces MG. $B$ and $V$ photometry for all targets was sourced from the catalogue which provides the smallest errors: either APASS DR9 (\citealt{2016a_Henden}), UCAC4 (\citealt{2013a_Zacharias}) or NOMAD (\citealt{2005a_Zacharias}). All $K_{\rm s}$ photometry is from 2MASS (\citealt{2003a_Cutri}).

\nocite{1999a_Voges}
\nocite{2000a_Voges}
\nocite{2006a_Pollacco}
\nocite{2007a_Norton}
\nocite{2016a_Boller}

\subsection{Selecting potentially young stars in SuperWASP}\label{SS_SuperWASP}

We learned from BJM15 that the majority of the raw lightcurves of young stars exhibited irregular features as well as sinusoidal behaviour, indicative of star-spot modulation and/or chromospheric plages -- both prognostic of stellar youth. We assigned priority to any objects whose lightcurves displayed any of these features, but also demanded clean lightcurves and consistent rotation period measurements over several seasons of observations. We also tried to screen out, through manual inspection of the lightcurves, targets that looked like eclipsing binaries.

\begin{figure*}
\begin{center}
\includegraphics[width=0.9\textwidth]{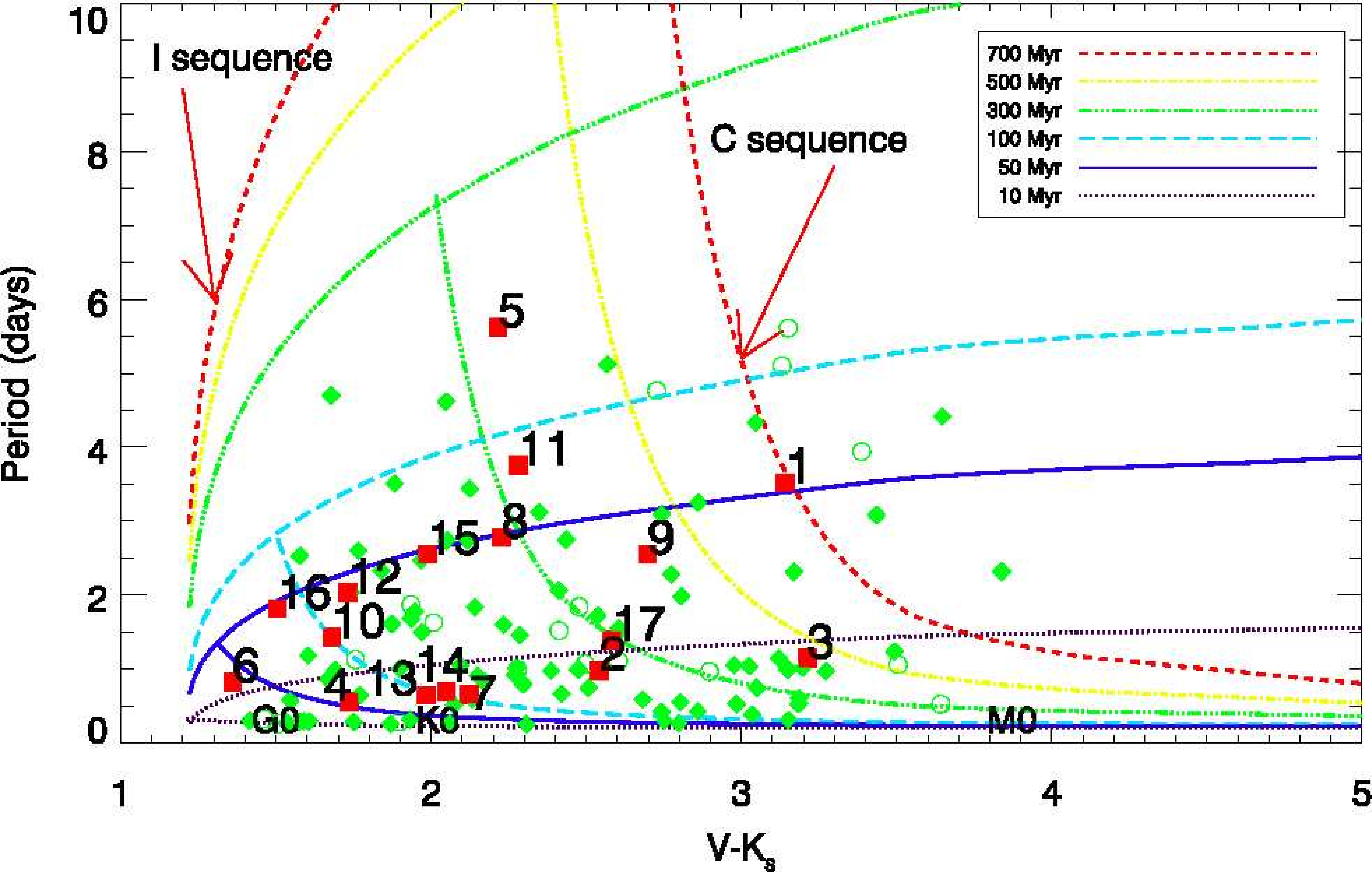}
\end{center}
\begin{flushleft}
 \caption{Rotation periods and \VK~colour for the entire sample observed in this paper. Red squares represent 17 objects that were identified
to be likely-young stars and were considered likely-single (referred to as the ‘likely-young sample’, see $\S$\ref{SS_LYS}). These objects are labelled from 1 to 17 and their corresponding SuperWASP names are provided in Table~\ref{T_PX_LYS}. Any objects that had two or more separate RV measurements whose mean values varied by more than 2 error bars from each other were assumed to be short-period binaries (see $\S$\ref{SS_Binary}) and are denoted here by open green circles. All other objects are shown with filled green diamonds. This symbol scheme is used throughout this work, unless otherwise stated. The various lines represent the I- and C-sequence gyrochrones proposed by \protect\cite{2003a_Barnes} and calibrated by \protect\cite{2008a_Mamajek}, plotted at a range of ages.}
 \label{F_Gyrochrones}
\end{flushleft}
\end{figure*}

The reader is referred to section 2 of BJM15 for a description of the SuperWASP project and post-processing techniques used to measure rotation periods, which are identical to those carried out in this work. Table~\ref{T_PX_LYS} shows the details of the period measurements for the sample of likely young and single objects in our sample, which are defined in $\S$\ref{SS_LYS}. Figure~\ref{F_Gyrochrones} displays rotation periods as a function of \VK~colour for our entire observed sample. Details for the other observed targets are provided online. We obtained spectroscopy for 122 stars, 103 selected from the criteria described here, and a further 19 which satisfy these criteria, but have proper-motions consistent with membership of the Pisces MG (see next section).

{\scriptsize
\begin{table*}
  \caption{Measured periods for the sample of 17 likely-young objects (see $\S$\ref{SS_LYS}). Periods are calculated (and standard errors) following the methodology described in section 2 of BJM15. The first error bar is the (averaged) uncertainty calculated using equation 2 in \protect\cite{2010a_Messina}. The second error bar (where appropriate) is the standard error in measurements for 2 or more seasons. The column `$\Delta\chi^{2}$' is indicative of the power difference between the peaks of the 2 strongest features in each periodogram and described in detail in BJM15, where $\Delta\chi^{2} > 1000$ is considered good (we note that targets 1 and 17 have $\Delta\chi^{2} < 1000$, however we retain them in our analysis). `$N$' refers to the number of seasonal lightcurves analysed for each target star. The column labelled `$Q$' refers to the quality of the period determination (described in section 2 of BJM15). X-ray count rates (CR) and HR1 ratios are extracted from either the 1RXS or 2RXP catalogs and $\log L_{X}/L_{\rm bol}$ is the X-ray to bolometric luminosity ratio (see $\S$\ref{SS_Xray}). $W1$ and $W4$ photometry from the Wide-field Infrared Survey Explorer (WISE) are given in columns 12 and 13, respectively. The equivalent data for all other objects observed in this work (and for all subsequent tabular data) are available online.}
\begin{center}
\begin{tabular}{p{0.1cm}p{2.5cm}p{0.5cm}p{0.9cm}p{2.8cm}p{0.6cm}p{0.2cm}p{0.2cm}p{0.9cm}p{0.6cm}p{1.4cm}p{0.7cm}p{0.7cm}}
\hline
\hline
\# & SuperWASP ID         & $V$       & $V-K_{\rm s}$ &  Period                               & $\Delta\chi^{2}$ & $N$ & $Q$ &                CR  &       HR1 & $\log L_{X}/L_{\rm bol}$ & $W1$ & $W4$ \\
      & (1SWASP J-)          & (mag)     & (mag)         &  (days)                               &                  &     &     & (cts\,s$^{-1}$) &           &                              & (mag) & (mag) \\
\hline
1  &   001712.81+023646.3    &     12.76 &      3.14 &     3.519 $\pm$     0.008 $\pm$     0.014 &           13 &       4 & A &      0.07 &   $+$0.13 &    $-$1.810 &  9.6 & 8.3 \\
2  &   010705.51+190908.3    &     10.11 &      2.54 &     0.978 $\pm$     0.017 $\pm$     0.563 &         2591 &       2 & C &      0.52 &   $-$0.07 &    $-$2.919 &  7.5 & 7.4 \\
3  &   013723.23+265712.1    &     10.85 &      3.21 &     1.141 $\pm$     0.017 $\pm$     0.074 &         1188 &       2 & C &      0.28 &   $-$0.04 &    $-$3.038 &  7.5 & 7.6 \\
4  &   033128.99+485928.3    &     10.19 &      1.74 &     0.558 $\pm$     0.022 $\pm$     0.195 &        26074 &       3 & C &      0.08 &   $+$1.00 &    $-$3.346 &  8.4 & 8.5 \\
5  &   155444.26$-$075142.8  &     11.38 &      2.22 &     5.626 $\pm$     0.020                 &        13988 &       2 & B &      0.04 &   $+$1.00 &    $-$2.222 & 13.0 & 8.4 \\
6  &   172011.53+495456.0    &     11.48 &      1.36 &     0.821 $\pm$     0.011 $\pm$     0.097 &         3995 &       8 & C &      0.03 &   $+$0.16 &    $-$2.648 & 10.1 & 8.9 \\
7  &   174121.23+084334.0    &     11.10 &      2.12 &     0.655 $\pm$     0.019 $\pm$     0.432 &        13528 &       2 & C &      0.07 &   $+$0.85 &    $-$3.062 & 12.1 & 8.7 \\
8  &   174624.31+222859.5    &     10.90 &      2.23 &     2.779 $\pm$     0.021 $\pm$     1.203 &         2907 &       3 & C &      0.10 &   $-$0.12 &    $-$2.844 &  8.6 & 8.9 \\
9  &   174746.93+521340.2    &     11.38 &      2.70 &     2.553 $\pm$     0.014 $\pm$     0.805 &         1102 &       7 & C &      0.06 &   $+$0.50 &    $-$2.837 &  8.6 & 8.3 \\
10 &   181631.34+284811.5    &      9.44 &      1.68 &     1.426 $\pm$     0.011                 &         3069 &       1 & C &      0.06 &   $-$0.34 &    $-$3.434 &  7.7 & 7.6 \\
11 &   215041.07+143413.7    &     11.16 &      2.28 &     3.752 $\pm$     0.012 $\pm$     0.055 &        18490 &       6 & A &      0.05 &   $+$0.00 &    $-$3.175 &  8.8 & 8.4 \\
12 &   223426.95+404232.3    &     11.03 &      1.73 &     2.028 $\pm$     0.014 $\pm$     0.928 &        58505 &       7 & C &      0.02 &   $+$1.00 &    $-$3.785 &  9.2 & 7.5 \\
13 &   225538.21+281012.5    &     11.63 &      1.98 &     0.648 $\pm$     0.009 $\pm$     0.269 &         9684 &       5 & C &      0.05 &   $+$0.41 &    $-$2.855 & 13.5 & 7.8 \\
14 &   232023.15+164723.3    &     11.02 &      2.05 &     0.686 $\pm$     0.021 $\pm$     0.001 &         7296 &       3 & A &      0.11 &   $+$0.43 &    $-$2.964 &  8.9 & 8.2 \\
15 &   232156.36+072132.8    &     10.92 &      1.99 &     2.550 $\pm$     0.014 $\pm$     0.012 &         4629 &       4 & A &      0.01 &   $+$0.17 &    $-$3.001 &  8.7 & 8.2 \\
16 &   234006.09$-$040255.2  &      9.91 &      1.50 &     1.812 $\pm$     0.020 $\pm$     0.011 &        18448 &       4 & A &      0.14 &   $+$0.15 &    $-$3.344 &  8.3 & 8.0 \\
17 &   234945.36+312627.4    &     12.83 &      2.58 &     1.389 $\pm$     0.010 $\pm$     0.009 &          741 &       3 & A &      0.06 &   $+$0.42 &    $-$3.275 & 10.2 & 9.0 \\
\hline
\end{tabular}
\end{center}
\label{T_PX_LYS}
\end{table*}
}

\subsection{Potential members of Pisces}\label{SS_Potential_Pisces}

To select stars likely to be co-moving with the Pisces MG we used the proper-motion selection method described in section 2 (equations 1--5) of \cite{2009a_Lepine}. The basic principle is to find the local projected motion of the MG in the plane of the sky ($v_{l_{\rm MG}}$, $v_{b_{\rm MG}}$, both functions of Galactic space velocities $UVW$, in which $U$ is the velocity in the direction towards the Galactic centre, $V$ in the direction of Galactic rotation and $W$ towards the Galactic North pole) in the direction of Galactic longitude $l$ and Galactic latitude $b$. We calculate $\Phi$, the angle between a star's local proper-motion vector ($\mu_{l_{\rm MG}}$, $\mu_{b_{\rm MG}}$) and the expected local projected motion if it were a MG member ($v_{l_{\rm MG}}$, $v_{b_{\rm MG}}$). In BJM15 we reported the velocity range of seven objects we considered to define the Pisces MG as $U = -10.5 \pm 2.7 (\pm 1.8)$; $V = -4.3 \pm 1.0 (\pm 1.4)$; $W = -4.9 \pm 3.6 (\pm 1.3)$\kms~(error bars are the standard deviations and the values in parentheses are the average $UVW$ uncertainties), therefore to test our objects for kinematic coherence we assigned dummy $U, V$ and $W$ velocities for each object in steps of 1.0\,\kms~between $-15.0 < U -6.0$; $-6.7 < V < -1.7$; $-9.8 < W < +0.2$, and if any combination of these $UVW$ corresponded to $\Phi < 2^{\circ}$ then it was retained as a candidate Pisces MG member.

A second criterion was based on the possible age of the Pisces MG (30--200 Myr) and on position in a colour-magnitude diagram. The ``presumed'' sets of successful $UVW$ coordinates for each star were used to estimate a kinematic distance ($d_{\rm kin}$, see equation 6 in \citealt{2009a_Lepine}), which can then be compared to a photometric distance, and we demanded that:
\\
\begin{equation}
|K_{\rm s} + 5\log(d_{\rm kin}) + 5 - M_{K_{\rm s}}(V-K_{\rm s})| < \sigma_{K_{\rm s}},
\end{equation}
\\
where $M_{K_{\rm s}}(V-K_{\rm s})$ is a fifth order polynomial fit to a main-sequence (1\,Gyr) isochrone in \cite{2015a_Baraffe}. Given that the latest spectral-type in the seven objects that define the Pisces group is K4 and that the difference in $K_{\rm s}$ magnitude between a mid-K star at 30\,Myr and on the main sequence is $\sim 0.5$\,mag (\citealt{2008a_Hillenbrand}), we demand $\sigma_{K_{\rm s}} < 1.5$ (for at least one combination of $UVW$ for each star),  which also accounts for photometry errors and binarity. 

Of the 160 stars in this list only 72 were observable during the telescope observing run and ten have previously published data, sufficient to measure a Galactic space velocity, however a literature search revealed 8 of these as members of the Taurus-Auriga star forming region and 2 as Pleiades members. We find that the proper-motion of the 10 Pisces candidates which turned out to be Taurus-Auriga/Pleiades objects point towards the Pisces convergent point, but have radial velocities (RVs) inconsistent with the Pisces MG, which is to be expected, since the $UVW$ centroid of the Pisces MG differs considerably from that of the Pleiades or Taurus-Auriga. This means that RVs are essential to separate young objects that might be members of the Pleiades or Taurus-Auriga from those that are part of the Pisces MG. In total, 19 Pisces candidates were observed on the telescope run. We note that TGAS data were not available when the targets were selected and observed.

\section{Observations}\label{S_Observations}

Observations were taken over 7 consecutive nights commencing $4^{\rm th}$ July 2016 using the 2.5-m Isaac Newton Telescope (INT) combined with the intermediate dispersion spectrograph (IDS). The H1800V grating, $4096 \times 4096$ pixel Red+2 CCD camera and a 1.4 arcsec slit gave a 2-pixel resolution of $0.7$\AA\ in the range $\lambda\lambda 6350-7020$\AA\ and a resolving power of $\sim 10^{4}$. Flat-fields, bias frames and sky exposures were taken at the beginning of each night and every observation was accompanied with an exposure of a CuNe+CuAr arc lamp (for a few observations the science frames were bracketed with an arc exposure either side of the observation), from which a wavelength solution was derived. Spectra were bias-corrected, flat-fielded, background-adjusted, trimmed, wavelength corrected and aperture extracted using standard procedures in {\sc IRAF}\footnote{IRAF is distributed by the National Optical Astronomy Observatories, which are operated by the Association of Universities for Research in Astronomy, Inc., under cooperative agreement with the National Science Foundation.}. No flux calibrations were applied to the spectra.

In order to calibrate heliocentric radial velocities (RVs) for our targets, observations were also made of several RV and low-activity template stars (selected for their extremely low levels of chromospheric Ca {\sc ii} H and K emission, characterised by $\log R_{\rm HK}$) over a similar spectral-type range as the targets. These were measured during twilight on each night and are listed in Table~\ref{T_RV_Standards}. Our method to obtain RVs is described in $\S$\ref{SS_RVs}.

\begin{figure}
 \begin{center}
\includegraphics[width=0.45\textwidth]{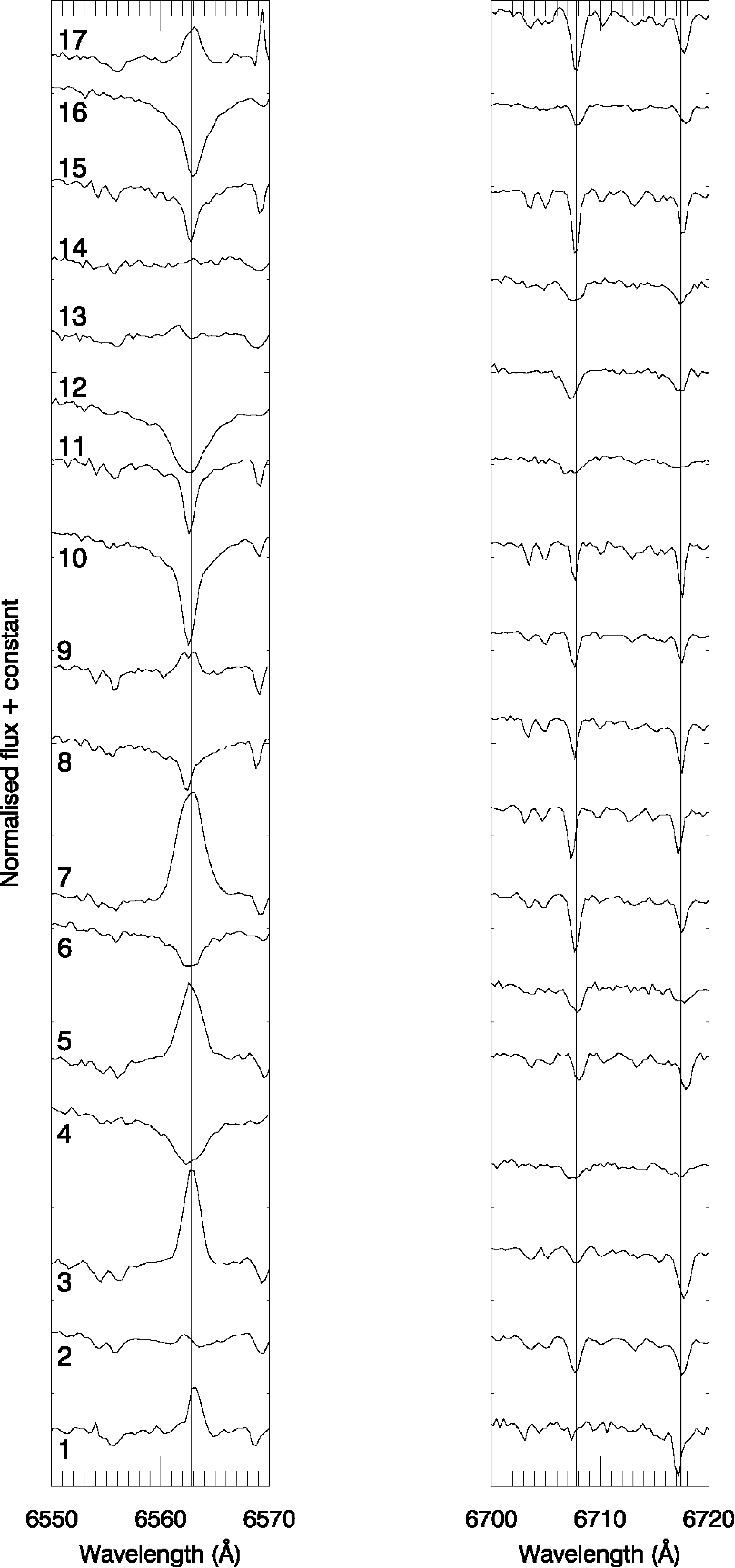}
 \end{center}
 \flushleft
 \caption{Heliocentrically corrected, normalised spectra for the 17 objects in the likely-young sample. The left panel highlights the H$\alpha$ feature around 6562.8\AA\ and the right panel displays the Li feature around 6707.8\AA.}
  \label{F_Spectra}
\end{figure}

The observing strategy followed the same protocols as BJM15: spectra were analysed in real-time directly after a science exposure and if there was an indication of a strong Li line at 6707.8\AA, the object was reobserved on a subsequent night to acquire a second RV measurement to test for binarity (see $\S$\ref{SS_Binary}). Figure~\ref{F_Spectra} shows the optical spectra for the 17 objects unlikely to be in tidally-locked short period binaries and $< 200\,$Myr (these are defined in $\S$\ref{SS_LYS}).

{\scriptsize
\begin{table}
  \caption{RV and minimum activity standards -- All RVs are from \protect\cite{2013a_Soubiran}, except for HD 114762 (\protect\citealt{2006a_Gontcharov}), HD 10780 (\protect\citealt{2002a_Nidever}), HD 4628 (\protect\citealt{2012a_Chubak}) and HD 160346 (\protect\citealt{2010a_Maldonado}). All $v\sin i$ values are from \protect\cite{2005a_Glebocki}. All $\log R′_{\rm HK}$ are from \protect\cite{2010a_Isaacson}, except for HD 114762 (\protect\citealt{2010a_Raghavan}) and HD 190007 (\protect\citealt{2011a_Mittag}).}
\begin{center}
\begin{tabular}{p{1.6cm}p{0.8cm}p{1.2cm}p{0.8cm}p{0.7cm}p{0.9cm}}
\hline
\hline
Standard  & SpT   & RV        & $\sigma$RV    & $v\sin i$ & $\log R′_{\rm HK}$ \\
          &       & \kms      & \kms   & \kms      & \\
\hline
HD 114762 & F8V   & 49.1      & 0.1   & 1.7        & $-4.902$ \\
HD 1461   & G0V   & $-10.086$ & 0.005 & $< 5.0$    & $-5.008$ \\
HD 10780  & K0V   & 2.64      & 0.07  & 0.9        & $-4.700$ \\
HD 4628   & K2.5V & $-10.229$ & 0.03  & 1.6        & $-4.979$ \\
HD 160346 & K3V   & 21.19     & 0.08  & 2.6        & $-4.956$ \\
HD 190007 & K4Vk  & $-30.227$ & 0.006 & 2.8        & $-4.592$ \\
HD 209290 & M0V   & 18.144    & 0.07  & 3.8        &          \\
\hline
\end{tabular}
\end{center}
\label{T_RV_Standards}
\end{table}
}

\section{Data reduction}\label{S_Data}

\subsection{Radial velocity measurements}\label{SS_RVs}

Each spectrum was split into 7 consecutive wavelength bins of 50\AA\ between 6400 and 6800\AA\, avoiding the region between 6550 and 6600\AA\ where the H$\alpha$ feature is likely to be ``contaminated'' by chromospheric emission. Using the {\sc IRAF} program $fxcor$, a pixel cross-correlation function (CCF) was calculated for each wavelength bin between a target spectrum and the spectrum of an RV standard (see Table~\ref{T_RV_Standards}). Radial velocity shifts (${\rm RV}_{\rm shift}$) were calculated by identifying the midpoint of a Gaussian fit to a maximum of 11 pixels either side of the CCF peak that are within the top 40 per cent of the CCF. 

For some target/template correlations $fxcor$ was unable to provide a good fit due to spectral-type mismatch, poor quality data and/or binarity and in some cases no CCF peak was measured at all. Therefore we removed any measurements that were $> 2\sigma$ from the mean value and reiterated the process a maximum of 5 times. A weighted average relative RV for each target was calculated over all bins for each template star using the $R$ quality factor in \cite{1979a_Tonry} and then transferred onto a heliocentric reference frame. In all cases the template that provided the lowest uncertainty in RV was used.

Three sources of error were considered and added in quadrature: the standard error measured over each wavelength bin (weighted by $R$), the published RV uncertainty of the RV standard used in the calibration as provided in Table~\ref{T_RV_Standards}, and a systematic uncertainty each night estimated from the cross-correlations of all the RV standards with one another. The RV measurements for each RV standard relative to one another formed a matrix of cross-correlation values. The systematic uncertainty on each night was estimated to be 0.3, 0.9, 2.1, 2.6, 1.1, 0.7, 0.8\,\kms, chronologically. The larger uncertainties on nights 3 and 4 are likely because the seeing was much smaller than the slit width. Generally, the dominant source of error in RV measurement was from the systematic uncertainty. RV measurements and errors are provided in Table~\ref{T_RVs_LYS}, along with any previous literature values. In some cases there are two or more RV measurements, either from our repeat observations or from the literature (or both). In these cases our final reported RV is the weighted mean.

{\tiny
\begin{table}
\caption{Radial velocity measurements for the likely-young sample. The Heliocentric Julian Date (HJD) in column 2 starts from 2457500 days. Individual RVs are provided in column 3 and the final averaged RV in column 4. The column titled `$B$' is the binary score based on the criteria described in $\S$\ref{SS_Binary}. The RV measurements for the rest of the targets can be found in the supporting online material.}
\begin{center}
\begin{tabular}{lrrrr}
\hline
\hline
Target                 & HJD        & $RV_{\rm indiv}$ & $RV_{\rm final}$ & $B$ \\
 
(SW)                   &  & \kms             & \kms             &     \\
\hline
0017+0236, 1           & 80.61      &    $5.0 \pm 1.1$ &    $5.0 \pm 1.1$ & 2   \\
0107+1909, 2           & 76.73      &    $5.7 \pm 1.0$ &    $4.6 \pm 2.4$ & 1   \\
                       & 77.63      &    $1.5 \pm 2.6$ &                  &     \\
0137+2657, 3$^{\rm a}$ & 77.69      &   $-8.7 \pm 2.6$ &   $-2.6 \pm 3.1$ & 1   \\
                       & 79.73      &    $0.1 \pm 0.8$ &                  &     \\
0331+4859, 4$^{\rm b}$ & 77.72      &    $0.3 \pm 2.7$ &   $-1.0 \pm 1.8$ & 1   \\
                       & 78.71      &   $-0.4 \pm 1.6$ &                  &     \\
1554$-$0751, 5         & 75.46      &   $22.1 \pm 2.1$ &   $22.1 \pm 2.1$ & 2   \\
1720$-$4954, 6         & 80.44      &  $-19.1 \pm 0.9$ &  $-19.1 \pm 0.9$ & 2   \\
1741+0843, 7           & 76.47      &  $-22.0 \pm 1.1$ &  $-22.0 \pm 1.9$ & 1   \\
                       & 77.48      &  $-22.0 \pm 2.6$ &                  &     \\
1746+2228, 8           & 77.49      &  $-26.0 \pm 2.6$ &  $-26.3 \pm 1.8$ & 1   \\
                       & 75.44      &  $-26.4 \pm 0.9$ &                  &     \\
1747+5213, 9           & 75.50      &  $-25.8 \pm 1.0$ &  $-26.2 \pm 1.9$ & 1   \\
                       & 77.42      &  $-27.4 \pm 2.6$ &                  &     \\
1816+2848, 10          & 75.57      &  $-25.9 \pm 1.0$ &  $-24.4 \pm 2.7$ & 1   \\
                       & 78.59      &  $-20.5 \pm 2.6$ &                  &     \\
2150+1434, 11          & 76.58      &   $-3.9 \pm 1.0$ &   $-3.8 \pm 1.8$ & 1   \\
                       & 77.59      &   $-3.5 \pm 2.6$ &                  &     \\
2234+4042, 12          & 78.61      &  $-12.4 \pm 2.7$ &  $-13.1 \pm 2.0$ & 1   \\
                       & 79.61      &  $-13.5 \pm 1.1$ &                  &     \\
2255+2810, 13          & 78.63      &    $0.4 \pm 1.2$ &   $-0.8 \pm 1.0$ & 1   \\
                       & 79.57      &   $-0.9 \pm 0.7$ &                  &     \\
                       & 80.57      &   $-1.4 \pm 0.9$ &                  &     \\
2320+1647, 14          & 75.69      &    $3.8 \pm 1.0$ &    $5.1 \pm 1.7$ & 1   \\
                       & 78.70      &    $6.8 \pm 1.3$ &                  &     \\
2321+0721, 15          & 75.72      &    $6.7 \pm 0.9$ &    $6.9 \pm 1.8$ & 1   \\
                       & 77.64      &    $7.4 \pm 2.6$ &                  &     \\
2340$-$0402, 16        & 75.70      &   $13.3 \pm 1.0$ &   $13.9 \pm 1.2$ & 1   \\
                       & 78.70      &   $14.6 \pm 1.2$ &                  &     \\
2349+3126, 17          & 79.60      &    $4.5 \pm 0.7$ &    $3.7 \pm 1.1$ & 1   \\
                       & 80.63      &    $2.8 \pm 0.9$ &                  &     \\
\hline
\end{tabular}
\end{center}
a: RV = $-5.9 \pm 3.0$ \kms \citep{2010a_Schlieder}, b: RV = $-1.80 \pm 0.87$ \kms \citep{2007a_White}.
\label{T_RVs_LYS}
\end{table}
}

\subsection{Identifying binary objects}\label{SS_Binary}

Following BJM15, we assign a ``binary score'' from 1 to 5 for each target based on the likelihood of an object being part of a tidally-locked short-period binary (TLSPB) system (see $\S$4.5 in BJM15)\footnote{At the $< 5$ day rotation periods of our selected targets it is extremely likely that tidal synchronisation has already occurred and that the measured period is both the rotation of the components and the orbital period of the binary  (e.g. \citealt{2010a_Torres}).}. The scoring system is based on considerations of RV changes and line profiles. For the targets that were observed twice, the timing of the second observation was chosen to avoid observing the target at the same rotational phase. Targets with a binary score of 1 are most likely to be single and 5 most likely to be TLSPBs. Usually TLSPBs have rotation periods on the order of a few days (\citealt{1989a_Zahn, 2006a_Meibom}), and therefore we suspect a significant number ({\it at least} 10 per cent based on our observations in BJM15) of our targets to be rapidly rotating due to their binarity, as opposed to being young. We have slightly modified our scoring criteria from BJM15; a score of 1 is given to objects with two or more RV measurements and errors less than 5\,\kms~that are {\it consistent to within 2 error bars of one another}, and a score of 5 for objects that have two or more RV measurements that do not agree within 2 error bars of one another. Objects categorised with a score of 1 or 2 are most likely to be single stars or at least in wide binaries where the binarity does not influence the rotation rate. From our 122 observed targets we find 23 that score 1, 59 with a score of 2, 13 that score 3, 8 scoring 4 and 19 with a binary score of 5. We provide the basic properties of the 19 (+14 objects from BJM15) very likely TLSPBs (that score 5) in Appendix B. 

It is possible that some of the objects that appear single, have single-peaked CCFs and/or do not show evidence for RV variations in multiple visits are still binary systems. In BJM15 we performed simulations which showed that, for {\it single lined} spectroscopic binaries, the probability that we would see $\Delta{\rm RV} < 5$\,\kms~between two observations taken at random phases of a TLSPB with the same orbital period as the measured rotation period was $< 10$ per cent. In this paper we improved on BJM15 in that we tried to make sure that repeated observations were taken at a different phase (based on the rotation period), which further reduces this probability, since in such cases we might expect the RV to vary by tens of \kms. 

The situation is a little more complex for binaries with approximately equal mass components since if they are observed at an orbital phase such that the two stars produce an unresolved single CCF peak, then the RV of that peak will reflect the systemic velocity of the binary and would not vary if the components were not resolved in a subsequent observation. The spectral resolution of our detector is  $\sim 32$~\kms, and equal-mass (q=1) binaries separated by more than this velocity would display resolved peaks in their CCF. However, if the RV {\it separation} changed by more than about half this, then there would be a detectable change of the width of the CCF, even if the peak is at the same RV. Inspection of the CCF profiles of all the stars with multiple RV measurements has not revealed any with significant changes in CCF width between observations. Using the same simulations as described in BJM15 we estimate that $\sim 10$ per cent of short period binaries have components that might appear unresolved in any particular spectrum and CCF. However, the chances of then also observing an unresolved CCF on a second visit and for the CCF to show no detectable change in width is less than 1 per cent. Thus whilst there is still a signifcant probability that stars we have observed only once are roughly equal-mass, unresolved TLSPBs, our simulations suggest we would identify a q=1 TLSPB with 2 CCF measurements $> 99$ per cent of the time.

\subsection{Measuring equivalent widths and abundances}\label{SS_EWs}

We followed the same procedure in BJM15 to measure equivalent widths (EWs) and uncertainties for the Li~{\sc i} 6707.8\AA\ line and corrected for a blended Fe line at 6707.4\AA\ by subtracting $20\times(B-V)-3$\,m\AA\ from the measured EW (\citealt{1993a_Soderblom}). Where no Li line is apparent, or where the Li EW is found to be less than 40m\AA\ we quote 2$\sigma$ upper limits. For targets with 2 or more Li EW measurements we adopt the weighted mean and standard error as our final value (previous Li EW measurements in the literature are only used for comparison because several do not have quoted errors), unless they are both upper limits in which case we quote their average 2$\sigma$ upper limit. There are no cases in which we have a mixture of measurements and upper limits for a target. No errors are calculated for H$\alpha$ EWs, but for strong lines (H$\alpha$ EW $\approx 0.5$\AA) errors are assumed to be 0.1\AA. When there are 2 or more spectra for a target we take the mean and standard deviation as our final H$\alpha$ EW measurement.

Li abundances and effective temperatures ($T_{\rm eff}$) are calculated using the same methods as BJM15, and for stars with $T_{\rm eff} > 4000\,$K we use the same curves of growth, interpolation tables and non-local thermodynamic equilibrium corrections (\citealt{1993a_Soderblom, 1994a_Carlsson}). For stars cooler than 4000\,K we use the curves of growth in \cite{2002a_Zapatero-Osorio}, but without non-local thermodynamic equilibrium corrections. We linearly interpolate spectral-types from \VK~using table 5 of \cite{2013a_Pecaut}, assuming there is no significant reddening for these relatively nearby stars and a precision of one spectral sub-class. The EWs, photometry, $T_{\rm eff}$, spectral-types and Li abundances are provided in Table~\ref{T_Temp_EW_Abun_LYS} along with previously measured EWs in the literature.

{\tiny
\begin{table*}
  \caption{Equivalent widths (EWs) for the Li~{\sc i} 6707.8\AA\ and H$\alpha$ lines, $BVK$ photometry, temperatures, spectral-types (SpT) and Li abundances.  Li EW$_{\rm c}$ is the final EW after correcting for the blended Fe~{\sc i} line. Corresponding measurements for all the other targets observed in this run are supplied in the supplementary online material.}
\begin{center}
\begin{tabular}{lrrrrrrrrr}
  \hline
  \hline
Target       & Li EW  & Li EW$_{\rm c}$ & H$\alpha$ EW & $B$                & $V$                & $K_{\rm s}$        & $T_{\rm eff}$ & SpT  & A(Li)                  \\
             & (m\AA) & (m\AA)          & (\AA)        & (mag)              & (mag)              & (mag)              & (K)           &      &                        \\
  \hline
1            & 114    &  $95 \pm 47$    & $-0.39$      & $13.844 \pm 0.057$ & $12.759 \pm 0.112$ &  $9.618 \pm 0.019$ & 4250          & K4   & $1.08^{+0.40}_{-0.30}$ \\
2            & 213    & $199 \pm 29$    & $-0.07$      & $10.980 \pm 0.011$ & $10.114 \pm 0.057$ &  $7.572 \pm 0.023$ & 4710          & K3   & $2.14^{+0.16}_{-0.16}$ \\
3$^{\rm a}$  & 86     &  $66 \pm 28$    & $-0.78$      & $12.027 \pm 0.053$ & $10.855 \pm 0.063$ &  $7.642 \pm 0.027$ & 4200          & K5   & $0.81^{+0.30}_{-0.22}$ \\
4$^{\rm b}$  & 144    & $135 \pm 22$    & $0.92$       & $10.799 \pm 0.059$ & $10.195 \pm 0.060$ &  $8.460 \pm 0.022$ & 5330          & G6   & $1.60^{+0.06}_{-0.06}$ \\
5            & 194    & $174 \pm 29$    & $-0.82$      & $12.545 \pm 0.072$ & $11.384 \pm 0.064$ &  $8.233 \pm 0.021$ & 4240          & K4   & $1.30^{+0.16}_{-0.15}$ \\
6            & 183    & $174 \pm 29$    & $1.65$       & $12.093 \pm 0.149$ & $11.482 \pm 0.109$ & $10.122 \pm 0.019$ & 6040          & G0   & $3.19^{+0.22}_{-0.21}$ \\
7            & 299    & $287 \pm 36$    & $-1.38$      & $11.870 \pm 0.025$ & $11.099 \pm 0.040$ &  $8.975 \pm 0.019$ & 5090          & G9   & $2.93^{+0.20}_{-0.21}$ \\
8            & 212    & $199 \pm 25$    & $0.34$       & $11.713 \pm 0.010$ & $10.904 \pm 0.010$ &  $8.678 \pm 0.014$ & 4990          & K0   & $2.43^{+0.12}_{-0.13}$ \\
9            & 146    & $126 \pm 20$    & $-0.23$      & $12.541 \pm 0.048$ & $11.382 \pm 0.076$ &  $8.684 \pm 0.026$ & 4580          & K3   & $1.67^{+0.15}_{-0.15}$ \\
10           & 142    & $133 \pm 18$    & $1.16$       & $10.044 \pm 0.121$ &  $9.444 \pm 0.046$ &  $7.764 \pm 0.026$ & 5590          & G4   & $2.66^{+0.13}_{-0.13}$ \\
11           & 123    & $108 \pm 17$    & $0.65$       & $12.050 \pm 0.081$ & $11.157 \pm 0.053$ &  $8.877 \pm 0.018$ & 4940          & K1   & $1.95^{+0.14}_{-0.13}$ \\
12           & 158    & $147 \pm 28$    & $1.21$       & $11.716 \pm 0.010$ & $11.029 \pm 0.039$ &  $9.297 \pm 0.020$ & 5530          & G5   & $2.68^{+0.16}_{-0.15}$ \\
13           & 231    & $220 \pm 39$    & $0.03$       & $12.347 \pm 0.071$ & $11.627 \pm 0.050$ &  $9.643 \pm 0.017$ & 5240          & G8   & $2.75^{+0.20}_{-0.20}$ \\
14           & 178    & $165 \pm 31$    & $-0.05$      & $11.798 \pm 0.019$ & $11.016 \pm 0.045$ &  $8.968 \pm 0.019$ & 5170          & G8   & $2.44^{+0.17}_{-0.16}$ \\
15$^{\rm c}$ & 277    & $264 \pm 37$    & $0.59$       & $11.717 \pm 0.028$ & $10.923 \pm 0.052$ &  $8.936 \pm 0.028$ & 5240          & G8   & $2.95^{+0.20}_{-0.21}$ \\
16           & 133    & $125 \pm 19$    & $1.28$       & $10.477 \pm 0.059$ &  $9.908 \pm 0.044$ &  $8.403 \pm 0.021$ & 5830          & G2   & $2.80^{+0.14}_{-0.13}$ \\
17           & 252    & $242 \pm 43$    & $-0.40$      & $13.5 \pm 0.3$     & $12.828 \pm 0.014$ & $10.245 \pm 0.022$ & 4680          & K2   & $2.30^{+0.20}_{-0.23}$ \\
\hline
\end{tabular}
\end{center}
\begin{flushleft}	
a: H$\alpha$ EW = $-1.41\,$\AA; Li EW = 90\,m\AA~\citep{2012a_McCarthy}, b: H$\alpha$ EW = $3.020\,$\AA; Li EW = $110\,$m\AA~\citep{2007a_White}, c: Li EW = $150\,$m\AA~\citep{2006a_Torres}. NB: No error for $B$ photometry could be sourced for target 17 (from the YB6 catalog, \citealt{2003a_Monet}), therefore we assume a conservative error bar of 0.3\,mag.
\end{flushleft}
\label{T_Temp_EW_Abun_LYS}
\end{table*}}

\section{Age determination}\label{S_Ages}

\subsection{Gyrochronology}\label{SS_Gyro}

Our search concentrated on solar-type stars because a larger fraction of rapidly rotating FGK stars will be younger than 100\,Myr compared to lower-mass stars (\citealt{1987a_Stauffer, 2003a_Barnes, 2009a_Irwin, 2010a_Messina, 2013a_Bouvier}). A potentially effective way to quantify this spectral-type dependency of rotational evolution is in the context of gyrochronology (Barnes et al. 2003; 2007), which attempts to provide an empirical relationship between rotation period, spectral-type and age, calibrated with data in open clusters. Gyrochronology defines two distinct sequences: the fast (C-) sequence on which stars are initially rapidly rotating, and eventually traverse onto the slow (I-) sequence. The bifurcation point is the point at which the C sequence meets the I sequence and the ``gyrochrones'' become bimodal redward of this point. Figure~\ref{F_Gyrochrones} displays the rotation periods and \VK~colours for our entire observed sample, and overplotted are sets of gyrochrones, which are the re-calibrated functional forms in \cite{2008a_Mamajek} and transformed to \VK~colours using table 5 in \cite{2013a_Pecaut}.

Ages are estimated by comparing the positions of stars with these gyrochrones (see BJM15 for the details). Column 2 in Table~\ref{T_Ages_LYS} provides the upper age limits calculated from gyrochronology for the likely-young sample. If the relationship between rotation period, colour and age was monotonic, we would have an excellent method to calculate stellar ages. However, in clusters aged 10--125\,Myr there are orders of magnitude spreads in rotation period at a given spectral type, which prevents a deterministic age (see, for example, \citealt{2013a_Gallet}). For this reason we use rotation periods and gyrochronology as a stellar youth indicator, but do not assign primacy to these ages in our final age estimates.

\nocite{2003a_Barnes}
\nocite{2007a_Barnes}

\subsection{X-ray activity}\label{SS_Xray}

Rapidly-rotating solar-type stars manifest enhanced levels of chromospheric and coronal activity. Active stars generate X-ray flux through powerful magnetic recombination events and magnetically heated coronae, which can be observed in X-ray surveys such as the ROSAT 1RXS, 2RXP and 2RXS catalogues (Voges et al. 1999, 2000; Boller et al. 2016). Our search criteria did not require any particular threshold level of X-ray emission -- an entry within the 1RXS, 2RXP or 2RXS catalogues was deemed sufficient. Count rates and hardness ratios were available for our entire observed sample (see columns 8 and 9 in Table~\ref{T_PX_LYS}), from which X-ray to bolometric luminosities (\LxLbol, column 10 in Table~\ref{T_PX_LYS}) were calculated using the formulae in \cite{1995a_Fleming} and Stelzer et al. (2001, see also section 5.3 in BJM15).

\nocite{1999a_Voges}
\nocite{2000a_Voges}
\nocite{2016a_Boller}
\nocite{2001a_Stelzer}

\begin{figure*}
    \begin{center}
            \includegraphics[width=0.8\textwidth]{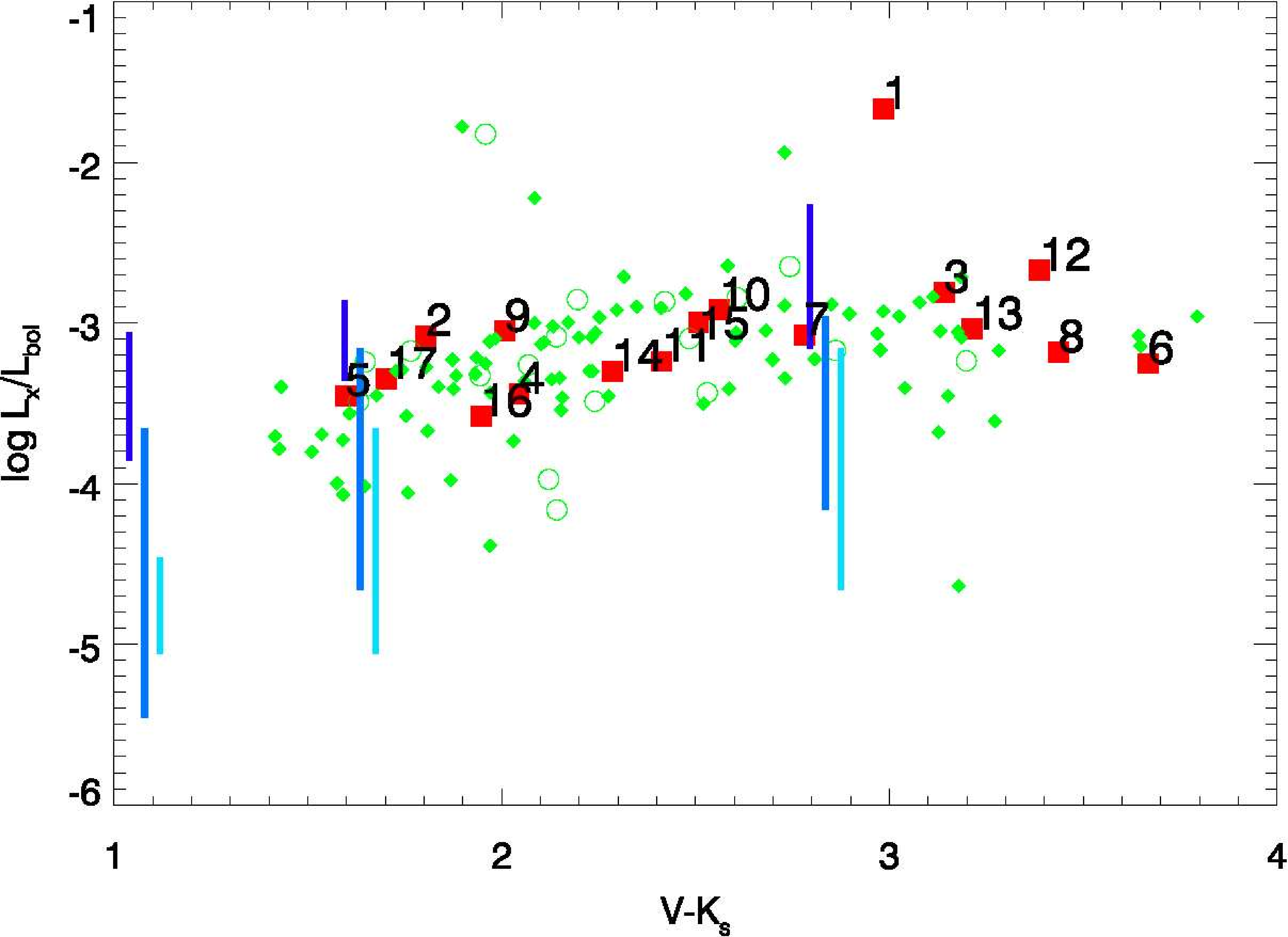}
    \end{center}
     \begin{flushleft}
\caption{$\log L_{\rm X}/L_{\rm bol}$ as a function of \VK. The $10^{\rm th}$ to $90^{\rm th}$ percentile range of \LxLbol~values in NGC~2547, the Pleiades and the Hyades are represented by the dark blue/azure/light blue lines, respectively.}
 \label{F_LX}
     \end{flushleft}
     \end{figure*}

Figure~\ref{F_LX} displays \LxLbol~vs \VK~for the observed sample. The dark blue, azure and sky blue lines represent the $10^{\rm th}$ to $90^{\rm th}$ percentile range in \LxLbol~for FGK stars in NGC~2547, the Pleiades and the Hyades ($\sim 35, 125$ and $625$\,Myr, respectively, based on figure 12 in \citealt{2006b_Jeffries}). All but one of our sample have coronal activity that is consistent with them being as young as the Pleiades or perhaps even younger -- in 5 cases \LxLbol~appears to be consistent with ages younger than NGC~2547. A problem with definitively assigning a very young age to these objects is that active stars do undergo X-ray flaring (see \citealt{2004a_Gudel,2016a_Tsuboi}). We do not find any conclusive evidence that the likely-young sample appears generally younger than the entire observed sample based on \LxLbol.

X-ray activity is not incorporated into our final age estimate as \LxLbol~is a relatively crude age indicator and largely rotation dependent in any case (\citealt{2014a_Jeffries}). The general trend between X-ray activity and age between 10 and 1000\,Myr in open clusters is shallow for FGK stars and a significant number of field stars have been observed with $\log L_{\rm X}/L_{\rm bol}$ variability greater than 1 dex (\citealt{2000a_Peres}; \citealt{2004a_Favata}; \citealt{2007a_Gudel}).

\subsection{H$\alpha$ measurements}\label{SS_H_alpha}

\begin{figure*}
 \vspace{2pt}
 \begin{center}
\includegraphics[width=0.8\textwidth]{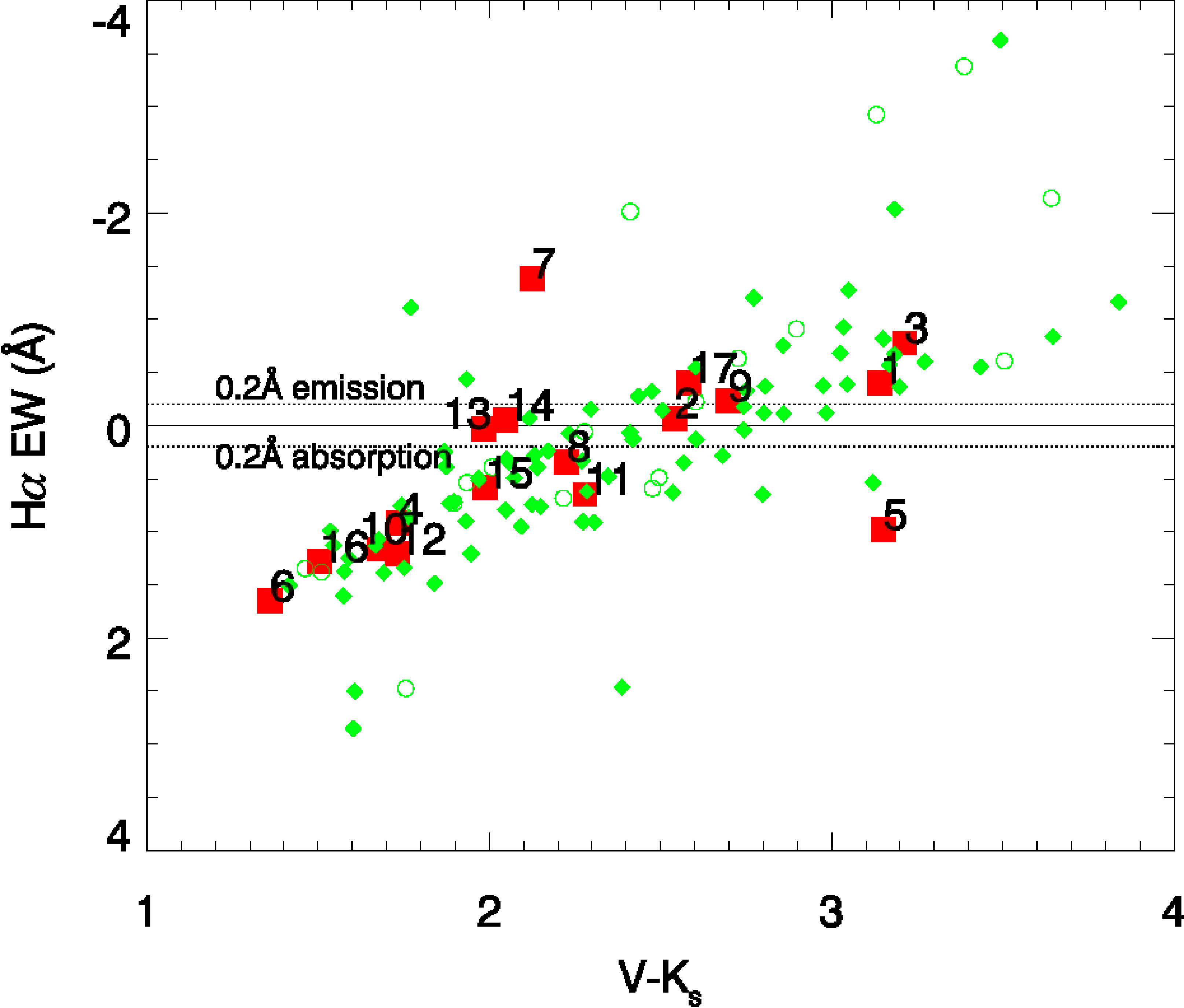}
 \end{center}
     \begin{flushleft}
 \caption{H$\alpha$ EW as a function of \VK. H$\alpha$ emission (negative EW) provides an upper age limit based on chromospheric activity. H$\alpha$ absorption provides a lower limit.}
      \label{F_Ha_EW}
    \end{flushleft}
\end{figure*}

Measuring the strength of the H$\alpha$ spectral feature, centered at 6562.8\AA\ provides a potentially more direct probe of stellar activity (\citealt{1999a_Hawley, 2003a_White}). It is well known that young, rapidly-rotating, active stars will exhibit strong H$\alpha$ emission which could therefore be an age indicator. Measurements of H$\alpha$ in several open clusters of known age show that low-mass stars remain chromospherically active and retain H$\alpha$ in emission for longer than solar-type stars (\citealt{1995a_Reid}). These provide empirical relations between H$\alpha$ emission/absorption, colour, and age -- see equations 5 and 6 in BJM15. Age upper limits are placed on stars with strong H$\alpha$ emission, and age lower limits for strong H$\alpha$ absorbers. It is difficult to discern whether stars with $|{\rm H}\alpha~{\rm EW}| < 0.2\,$\AA\ are in emission or absorption and their H$\alpha$ ages are given by a range defined by equations 5 and 6 in BJM15. The attained precision is somewhat misleading because the variability in H$\alpha$ EW emission/absorption can be $\sim 1.0\,$\AA\ in K-type stars in young clusters (see figure 8 in \citealt{1997a_Stauffer}). We therefore prescribe an additional 30\,Myr systematic uncertainty on all H$\alpha$ age limits quoted. The H$\alpha$ EWs are plotted against colour in Figure~\ref{F_Ha_EW} and the empirically determined H$\alpha$ ages are given in Table~\ref{T_Ages_LYS}.

\subsection{Comparing Lithium EW with clusters of known age}\label{SS_Li}

At present, evolutionary models are unable to fully capture the process of PMS Li depletion and are yet to explain the spread of $1-2$\,dex in Li abundance amongst K-type stars in the Pleiades (\citealt{1993a_Soderblom}), although a connection between rotation and radius inflation on the PMS has recently been suggested (Somers et al. 2015, 2017). Even if models are unable to predict the correct amount of Li depletion, an age can still be estimated by comparing Li measurements with those in open clusters of known age. As in BJM15, we estimate ages by comparing our target Li EWs as a function of \VK~(or equivalently, Li abundance versus $T_{\rm eff}$), with stars observed in the Hyades and Pleiades clusters, the $\gamma$~Velorum cluster, IC~2602 (references in BJM15). These clusters are de-reddened, respectively by $E(V-K_{\rm s})$ = 0.05, 0.11, 0.11 and 0.11\,mag, respectively (from the same references in BJM15). The targets observed in this paper are assumed to have small reddening.

\begin{figure*}
    \begin{center}
            \centering
            \includegraphics[width=0.8\textwidth]{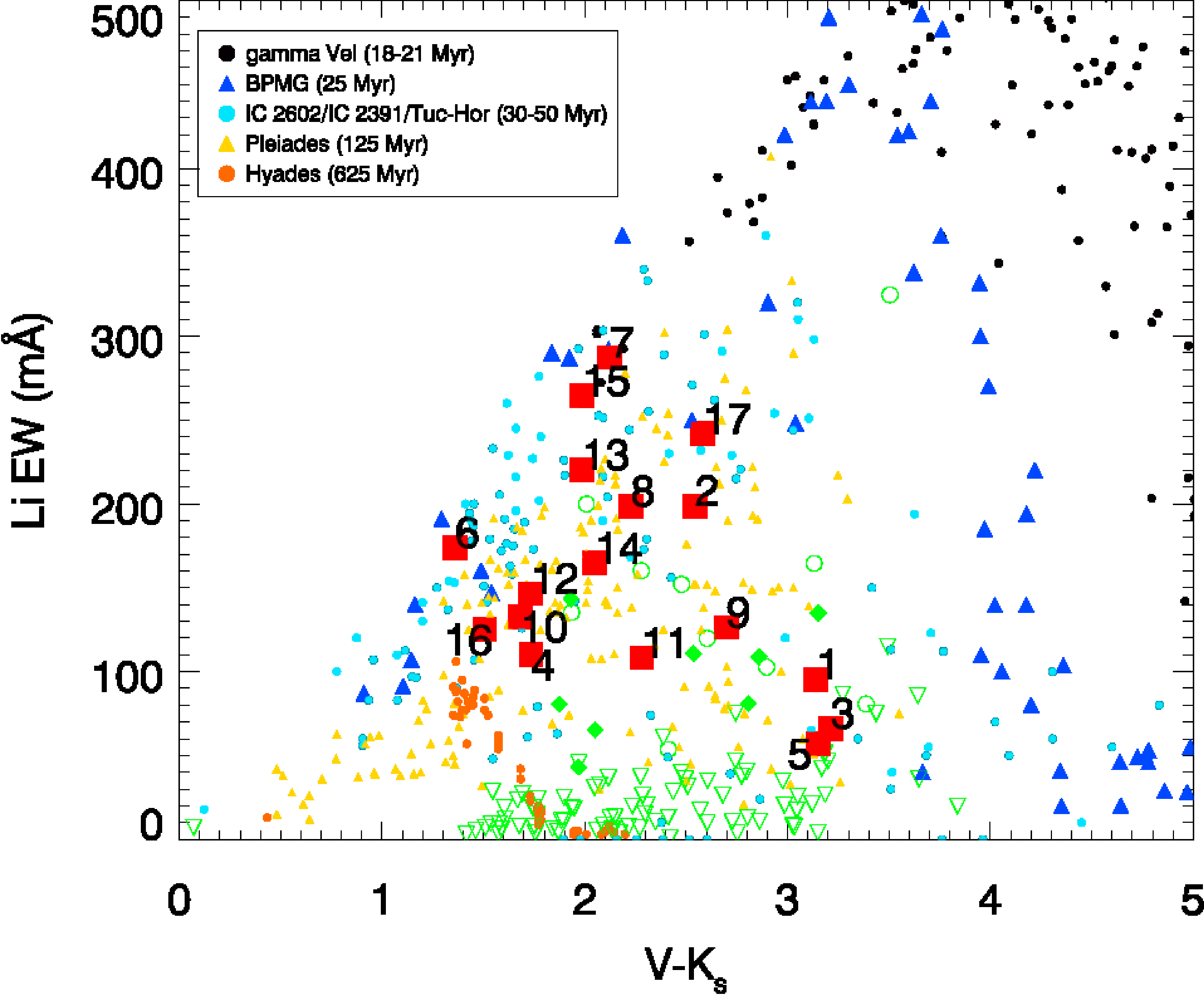}
    \end{center}
        \begin{flushleft}
 \caption{Li EWs and corresponding \VK~colours. Red squares, green open circles and green filled diamonds denote the same objects described in Figure~\ref{F_Gyrochrones}. Downward facing triangles represent $2\sigma$ upper limits. Objects are compared to members in the Hyades (625\,Myr), the Pleiades (125\,Myr), IC~2602 (30\,Myr), BPMG (25\,Myr) and $\gamma$~Vel (20\,Myr, see text for references) to estimate an Li-based age range. Error bars are not included, but are provided in Table~\ref{T_Temp_EW_Abun_LYS}.}
      \label{F_Li_EW}
    \end{flushleft}
\end{figure*}

\begin{figure*}
    \begin{center}
            \centering
            \includegraphics[width=0.8\textwidth]{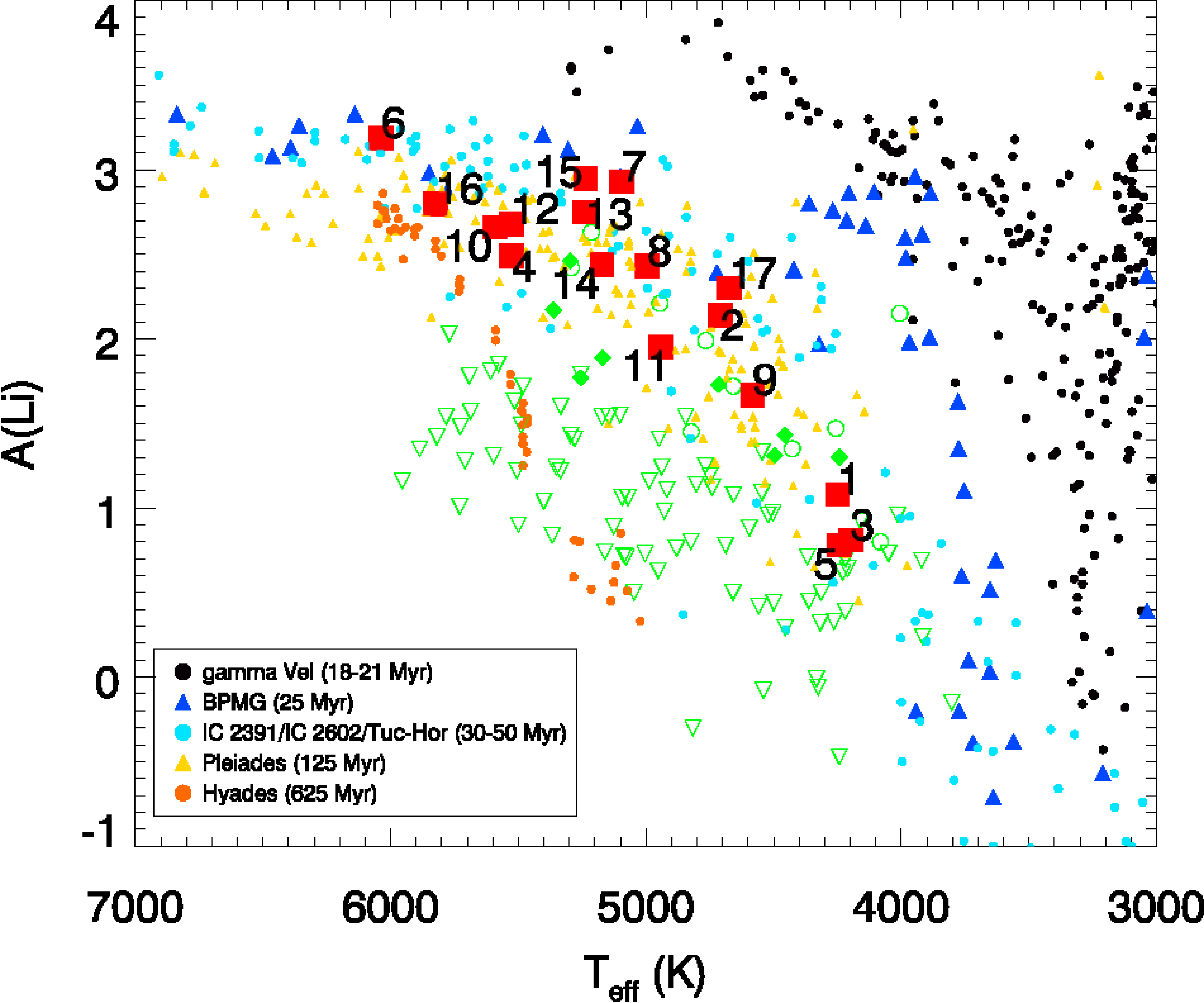}
    \end{center}
        \begin{flushleft}
 \caption{Li abundances as a function of surface temperature. Both the target sample and all ancillary data from the open clusters are folded through the same curve of growth (described in $\S$\ref{SS_EWs}). The symbol scheme is identical to Figure~\ref{F_Li_EW}.}
      \label{F_ALi}
    \end{flushleft}
\end{figure*}

In this work we also include Li EW data for the $\beta$ Pic MG (herein BPMG, age $\sim 20-25\,$Myr, \citealt{2014a_Binks, 2015a_Bell}) compiled from the lists in \cite{2016a_Messina} and the Tucanae-Horologium MG ($41 \pm 2\,$Myr, \citealt{2014a_Kraus}). We assume zero reddening for these nearby ($< 100$\,pc) groups. The age used for the $\gamma$~Velorum cluster has been revised to $18-21$\,Myr, following the work of \cite{2017a_Jeffries}. We display the Li EWs as a function of intrinsic \VK~for all of our targets in this paper, and for the fiducial clusters in the top panel of Figure~\ref{F_Li_EW} and their corresponding Li abundance versus $T_{\rm eff}$ in Figure~\ref{F_ALi}. Ages estimated from the Li measurements are given in column 4 of Table~\ref{T_Ages_LYS}. We note that a small amount of reddening would cause little difference to the age estimates.

We find, on visual inspection of the top panel in Figure~\ref{F_Li_EW}, that our young sample have Li EW ages definitely older than $\gamma$~Velorum. Targets 6, 7 and 15 could just be consistent with a BPMG age, but look more consistent with IC~2602 or even the upper envelope of the Pleiades distribution and are assigned a conservative age range of $25-150$\,Myr. Targets 4, 9 and 11 appear more consistent with the Pleiades (and not younger) and are estimated to be 100--200\,Myr, and the remaining stars in the likely-young sample appear to have Li EWs somewhere in the range covered by stars in IC~2602 and the Pleiades (30--200\,Myr).

\nocite{2015a_Somers}
\nocite{2017a_Somers}
\nocite{1990a_Soderblom}
\nocite{1990a_Soderblom}
\nocite{1990a_Soderblom}
\nocite{1990a_Soderblom}
\nocite{1990a_Soderblom}
\nocite{1990a_Soderblom}
\nocite{1990a_Soderblom}

\subsection{Colour-magnitude diagram}\label{SS_CMD}

A major development since the publication of BJM15 has been the release of positions, proper-motions and trigonometric parallaxes from the TGAS catalogue (\citealt{2016a_GAIA_Collaboration}). Trigonometric parallaxes, combined with photometry in two appropriate bands provide an age measurement by comparing the position of a star in a colour-magnitude diagram (CMD) with CMD positions at given ages predicted by evolutionary models (isochrones). We compiled trigonometric parallaxes from the literature, including the new TGAS catalogue, and for each target we adopted the parallax providing the smallest uncertainties (if 2 or more separate measurements were available). We found 80 objects in our spectroscopically observed sample with trigonometric parallaxes, all of which had their best measurements in the TGAS catalogue and thirteen of these are part of the likely-young sample defined in $\S$\ref{SS_LYS}. We also collected trigonometric parallaxes for the first time from TGAS for 15 of the likely-young sample from BJM15.

For objects with trigonometric parallaxes we calculate \MV~and corresponding uncertainties via a Monte Carlo process, where the mean and standard deviation in \MV~is derived from 10,000 iterations of normally distributed errors in $\pi$ (including the recommended 0.3\,mas systematic error) and $V$. For targets without a trigonometric parallax we used the same procedure as in BJM15: using their age range we estimate a maximum and minimum absolute $V$ magnitude from the isochrones of \cite{2015a_Baraffe}.

\begin{figure*}
    \begin{center}
            \centering
            \includegraphics[width=0.8\textwidth]{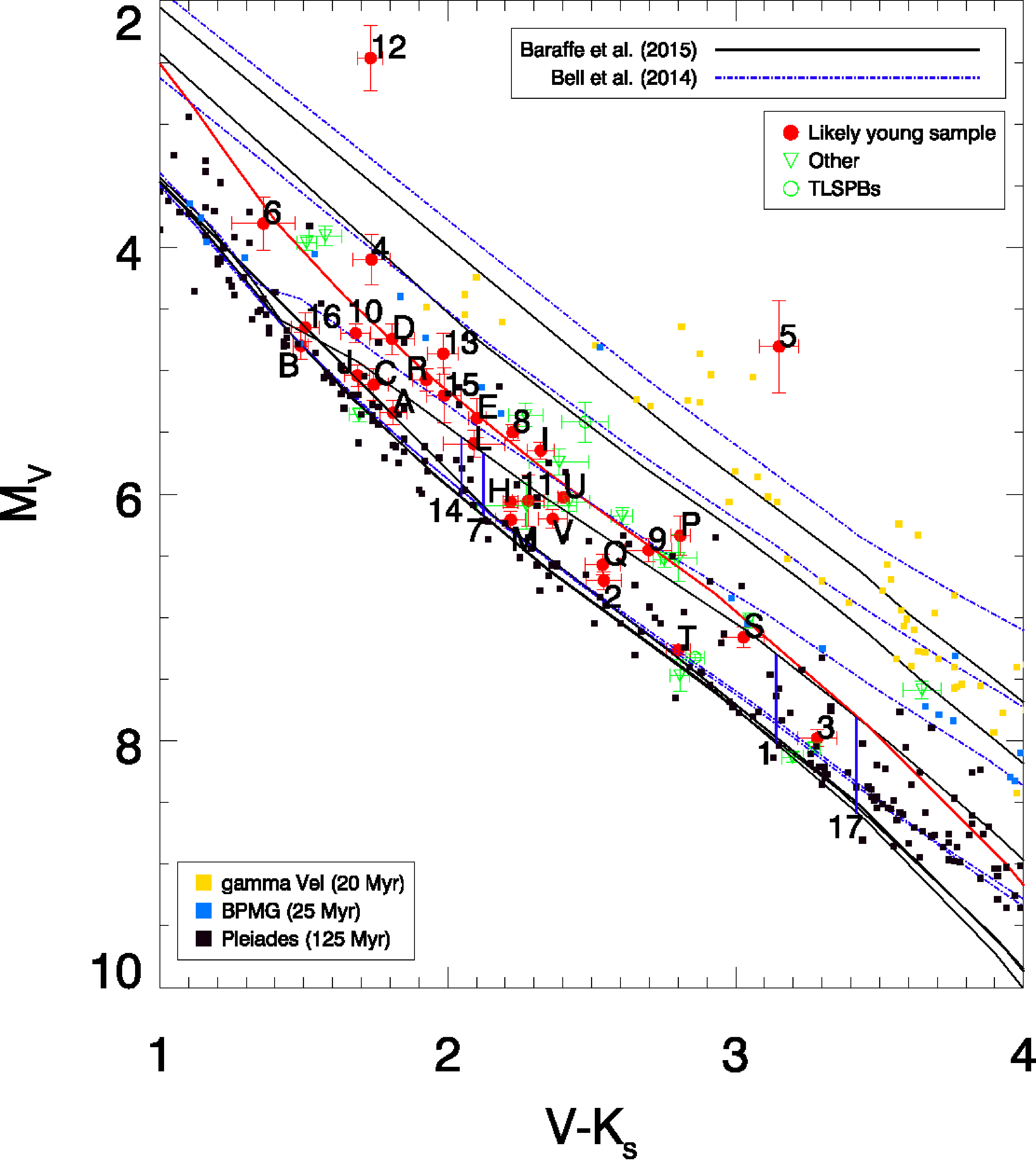}
    \end{center}
        \begin{flushleft}
 \caption{Absolute $V$ magnitude versus \VK~for objects with trigonometric parallaxes in the likely-young sample in this work and in BJM15 (both represented by filled red circles). Objects likely to be TLSPBs are represented by open green circles and stars unlikely to be young by green downwards-facing triangles. Blue dotted and black solid lines represent the \protect\cite{2014a_Bell} and \protect\cite{2015a_Baraffe} isochrones for ages 5, 10, 30, 100 and 200\,Myr. The solid red line represents a 1\,Gyr \protect\cite{2015a_Baraffe} isochrone shifted upwards by 0.75\,mag to represent main-sequnce equal-mass binaries. For young stars with no trigonometric parallax (identified in this work) a blue line provides the range of \MV, inferred from the age range calculated in $\S$\ref{SS_LYS} using the isochrones from \protect\cite{2015a_Baraffe}.}
      \label{F_CMD}
    \end{flushleft}
\end{figure*}

We plot the likely-young sample from both this work and those in BJM15 on an \MV~vs \VK~CMD in Figure~\ref{F_CMD} as filled red circles, with corresponding labels. Green open circles and green downwards-facing triangles represent TLSPBs or other objects in our spectroscopically observed sample, respectively, that have parallax errors $< 10$ per cent  (therefore avoiding Lutz-Kelker bias - Lutz $\&$ Kelker (1973); Oudmaijer et al. (1998)) and do not appear young based on their Li content. Overplotted as dotted blue and solid black lines are isochrones from \cite{2014a_Bell} and \cite{2015a_Baraffe}, respectively, at 5, 10, 30, 100 and 200\,Myr. For comparison we have included members of $\gamma$~Velorum (\citealt{2016a_Prisinzano}), BPMG (\citealt{2016a_Messina}) and Pleiades (\citealt{2016a_Rebull}).

\nocite{1973a_Lutz}
\nocite{1998a_Oudmaijer}

Several objects that are unlikely to be young based on their Li appear $1-2\,$mag above the main sequence. These objects may turn out to be unresolved binaries, which would place them above a single-sequence isochrone by as much as 0.75\,mag and partially explain their overluminosity. We plot a 1\,Gyr \cite{2015a_Baraffe} isochrone shifted upwards by 0.75\,mag to indicate the main-sequence equal-mass binary line. We expect our sample to be rapidly-rotating either due to being young, or being in short-period binary systems, therefore we expect a much larger binary fraction in our sample than a typical field population. The majority of TLSPB candidates uncovered in this work (those with binary scores of 5) are genuinely young (as measured by their Li EW) and should lie above the main-sequence.

For stars unlikely to be young that lie above the equal-mass binary line, further explanations are required for their CMD position. The targets are not luminous enough to be giants, but they could be subgiants. Because of the X-ray selection, our sample is an X-ray flux-limited sample. Such samples do contain a high fraction of rarer, but more luminous X-ray objects, such as RS~CVn stars -- active binaries with an evolved component or components (Strassmeier et al. 1988; Pandey \& Singh 2012). Whilst RS~CVn stars are more luminous X-ray sources, they are not expected to be young and Li-rich.

\nocite{1988a_Strassmeier}
\nocite{2012a_Pandey}

Since some of these stars have trigonometric distances beyond 100\,pc, significant reddening could cause a shift in the CMD. We tested this by comparing our objects with the FGK-stars in the 25\,pc volume-limited sample in \cite{1991a_Gliese} in a \VK~vs~$J-K_{\rm s}$~colour-colour diagram. None of our targets have a metric Euclidean separation of $> 0.2$ magnitudes from the loci of the \cite{1991a_Gliese} sample, suggesting that reddening does not account for their anomalous CMD positions.

Age ranges based on CMD position for the likely-young sample are provided in column 3 of Table~\ref{T_Ages_LYS}, where the age range is taken from the two \cite{2015a_Baraffe} isochrones closest to the target star in terms of $M_{\rm V}$ in Figure~\ref{F_CMD}. Targets 5 and 12 appear $> 1$ magnitude above the 5\,Myr isochrone, however neither appear younger than 30\,Myr based on their Li. We cannot suggest a simple explanation for these, however we speculate that the some of the factors contributing to the overluminosity of stars unlikely to be young may also cause targets 5 and 12 to appear much younger. Ages inferred from the CMD position of low-mass stars using the \cite{2015a_Baraffe} isochrones (or other models that do not consider the effects of magnetic fields and starspots) may be underestimated by as much as a factor of 2 for clusters $< 20$\,Myr (\citealt{2015a_Herczeg, 2016a_Feiden, 2017a_Jeffries}), but there may be smaller systematic underestimates even in older PMS stars.

In general the absolute magnitudes are too uncertain to assign an accurate age to stars unless they are substantially above the ZAMS. For the types of star we are considering this means younger than 50\,Myr even for K-type stars. We do not elect to incorporate CMD ages into our final age estimate because of the issues regarding accurate placement of \MV, \VK~and isochrones, however it can be used as supporting evidence for stellar youth. It is clear from Figure~\ref{F_CMD} that the likely young sample do appear to define a PMS population, and it seems likely that many of the objects that are not young, but not yet confirmed as binaries, are in fact binaries. Although we think the likely young sample are not in short period binaries, some of them could be in long period (but still unresolved) binaries and therefore up to 0.75\,mag more luminous than a single star -- probably in the same fraction as field stars -- which is a reason for not necessarily trusting the ages of individuals.

\subsection{Identifying mid-infrared excess}\label{SSS_IR_Excess}

Debris discs containing $\mu$m-sized particles form around 10\,Myr after star formation. Observations in young clusters indicate that the fraction of FGK stars with significant IR excess decreases from $\sim 40$ per cent at 10\,Myr to $\sim 10$ per cent by 100\,Myr (\citealt{2008a_Wyatt, 2011a_Zuckerman, 2017a_Meng}). While debris discs around field stars are rare, they are relatively common around stars younger than 100\,Myr.Therefore the presence of a debris disc may point towards stellar youth, but cannot determine youth unequivocally.

IR-based surveys such as the Wide-field Infrared Survey Explorer (WISE) have successfully identified hundreds of discs in the wavelength range $W1 = 3.6$ to $W4 = 22\mu$m. In Binks $\&$ Jeffries (2017) the scatter in $W1-W4$ colour around zero for disc-less FGK field stars was found to be $\sim 0.3$. Therefore in this work we searched the WISE database and flagged any objects with $W1-W4 > 1.0$ as potentially young from their IR-excess. In total we find $\sim 35$ per cent (6/17) of the likely-young sample have $W1-W4 > 1.0$ (as marked in Table~\ref{T_Ages_LYS}), however this fraction is $\sim 25$ per cent (26/105, possibly due to the presence of low-mass companions) for the remaining targets, therefore we do not advocate IR-excess as a discerning age indicator and only use it as supplementary supporting evidence of stellar youth. The $W1$ and $W4$ magnitudes are provided in Table~\ref{T_PX_LYS}.

\nocite{2017a_Binks}

\subsection{Combined age estimates and the likely-young sample}\label{SS_LYS}

Considering all the methods to obtain stellar ages in this section, we find that lithium ($\S$\ref{SS_Li}) is the only age-dating method capable of providing a strong constraint (both an upper and lower limit) on age throughout our entire sample. We therefore elect to use {\it lithium as our primary age indicator}, and utilise all other age indicators as supporting evidence.

From herein, we select a group of 17 stars for study, which are likely to be young ($25-200$\,Myr), evidenced by their photospheric lithium and are unlikely to be part of TLSPBs, with binary scores of 1 or 2 (see $\S$\ref{SS_Binary}), which we term the `likely-young sample'. Table~\ref{T_Ages_LYS} provides age estimates based on gyrochronology, H$\alpha$ and Li EWs. With the exception of targets 5 and 12 whose CMD ages are considerably younger than their Li ages, none of our age estimates for the likely-young sample are contradictory, however, all additional age indicators act only as {\it supporting evidence} for the targets being young. In general we find that the CMD ages are younger than the Li ages, consistent with some other recent work in young open clusters (e.g. \citealt{2016a_Feiden, 2017a_Jeffries, 2017a_Kastner}).

{\centering
\begin{table}
  \caption{Age estimates in Myr for the likely-young sample, based on Gyrochronology, H$\alpha$, the CMD and Li EW. Asterisks denote objects that have $W1-W4 > 1.0$. The final age estimate is solely from the Li EW age, other age indicators are used only as supporting evidence for the Li age.}
\begin{center}
\begin{tabular}{p{0.6cm}p{1.2cm}p{0.8cm}p{1.2cm}p{1.0cm}p{1.2cm}}
\hline
\hline
Label    & Target      & Gyro    & CMD       & H$\alpha$ EW & Li EW     \\
         & (SW)        & (Myr)   & (Myr)     & (Myr)        & (Myr)     \\
\hline
1$^{*}$  & 0017+0236   & $< 700$ &           & $< 83$       & $30-200$  \\
2        & 0107+1909   & $< 300$ & $100-200$ & $40-60$      & $30-200$  \\
3        & 0137+2657   & $< 500$ & $100-200$ & $< 91$       & $30-200$  \\
4        & 0331+4859   & $< 100$ &   $10-30$ & $> 20$       & $100-200$ \\
5$^{*}$  & 1554$-$0751 & $< 300$ &     $< 5$ & $> 84$       & $30-200$  \\
6        & 1720+4954   & $< 100$ &  $30-100$ & $> 12$       & $25-150$  \\
7$^{*}$  & 1741+0843   & $< 300$ &           & $> 26$       & $25-150$  \\
8        & 1746+2228   & $< 300$ &   $10-30$ & $> 37$       & $30-200$  \\
9        & 1747+5213   & $< 500$ &   $10-30$ & $> 47$       & $100-200$ \\
10       & 1816+2848   & $< 100$ &   $10-30$ & $> 19$       & $30-200$  \\
11       & 2150+1434   & $< 500$ &  $30-100$ & $> 40$       & $100-200$ \\
12$^{*}$ & 2234+4042   & $< 300$ &     $< 5$ & $> 20$       & $30-200$  \\
13$^{*}$ & 2255+2810   & $< 100$ &   $10-30$ & $20-30$      & $30-200$  \\
14       & 2320+1647   & $< 100$ &           & $20-30$      & $30-200$  \\
15       & 2321+0721   & $< 100$ &   $10-30$ & $> 41$       & $25-150$  \\
16       & 2340$-$0402 & $< 100$ &  $30-100$ & $> 15$       & $30-200$  \\
17$^{*}$ & 2349+3126   & $< 300$ &           & $< 41$       & $30-200$  \\
\hline
\end{tabular}
\end{center}
\label{T_Ages_LYS}
\end{table}}

\section{Kinematics}\label{S_Kinematics}

In this work the $UVW$ Galactic space velocities refer to $U$ in the direction of the Galactic centre, $V$ in the direction of Galactic rotation and $W$ towards the Galactic North pole. $X, Y$ and $Z$ are distances in the directions of $U, V$ and $W$, respectively. To calculate $UVW$ requires a measurement of position (right-ascension $\alpha$ and declination $\delta$), proper-motion ($\mu_{\alpha}$ and $\mu_{\delta}$), RV and parallax ($\pi$). Positions, calculated in the ICRS coordinate system were available for the entire observed sample in both the 2MASS catalogue (\citealt{2003a_Cutri}), and the Gaia first data release (TGAS, \citealt{2016a_GAIA_Collaboration}). Proper-motions were obtained from either the PPMXL (\citealt{2010a_Roeser}), UCAC4 or TGAS catalogue and are provided in columns 2 and 3 of Table~\ref{T_Kin_LYS}. For the likely-young sample, all positions are from TGAS, 4 objects have proper motions from UCAC4, 8 from PPMXL and 5 from TGAS. Radial velocities are provided in Table~\ref{T_RVs_LYS} and our method of sourcing parallaxes is described in $\S$\ref{SS_CMD}.

\begin{figure*}
    \begin{center}
            \centering
            \includegraphics[width=0.9\textwidth]{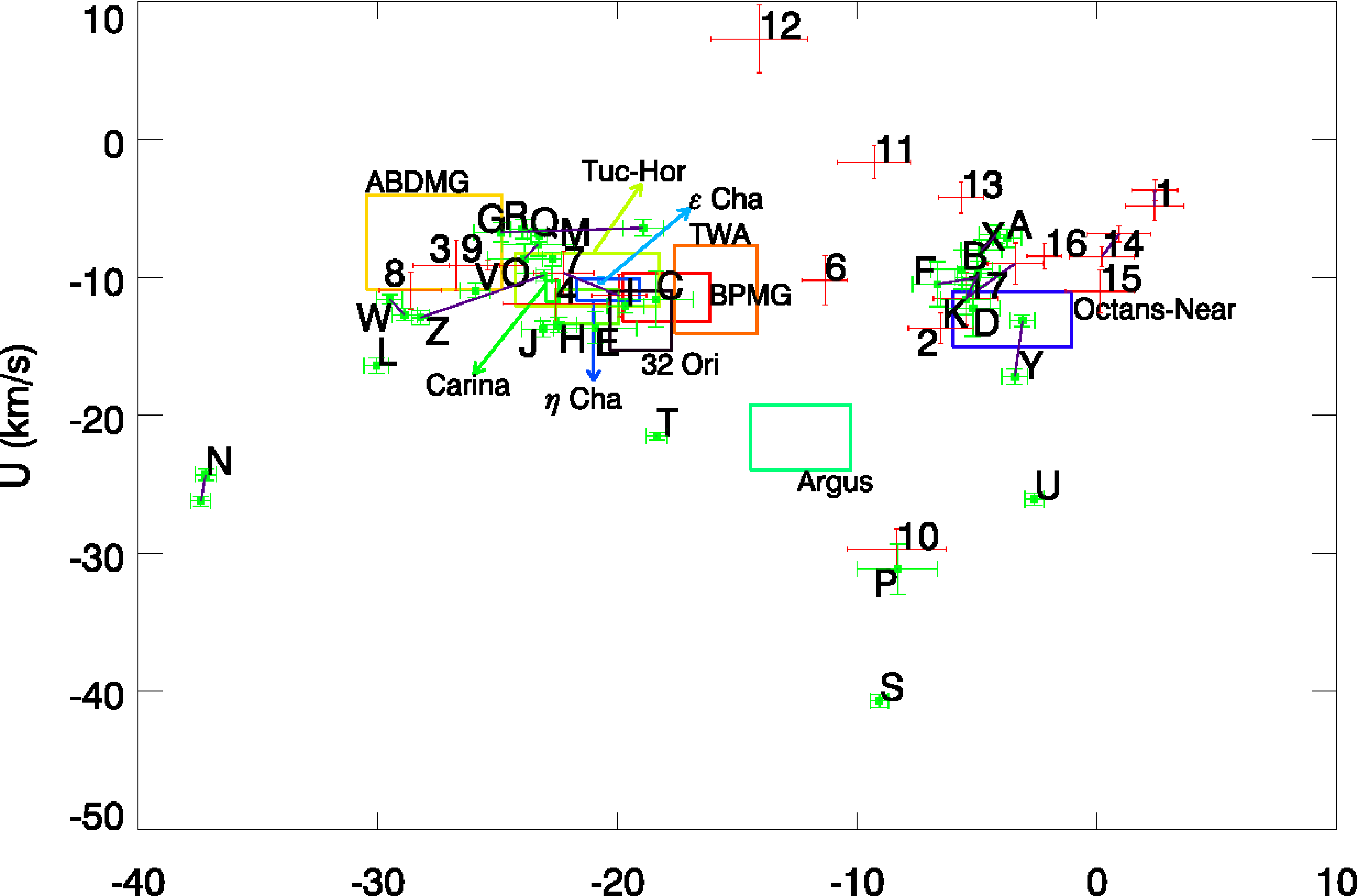}
            \includegraphics[width=0.9\textwidth]{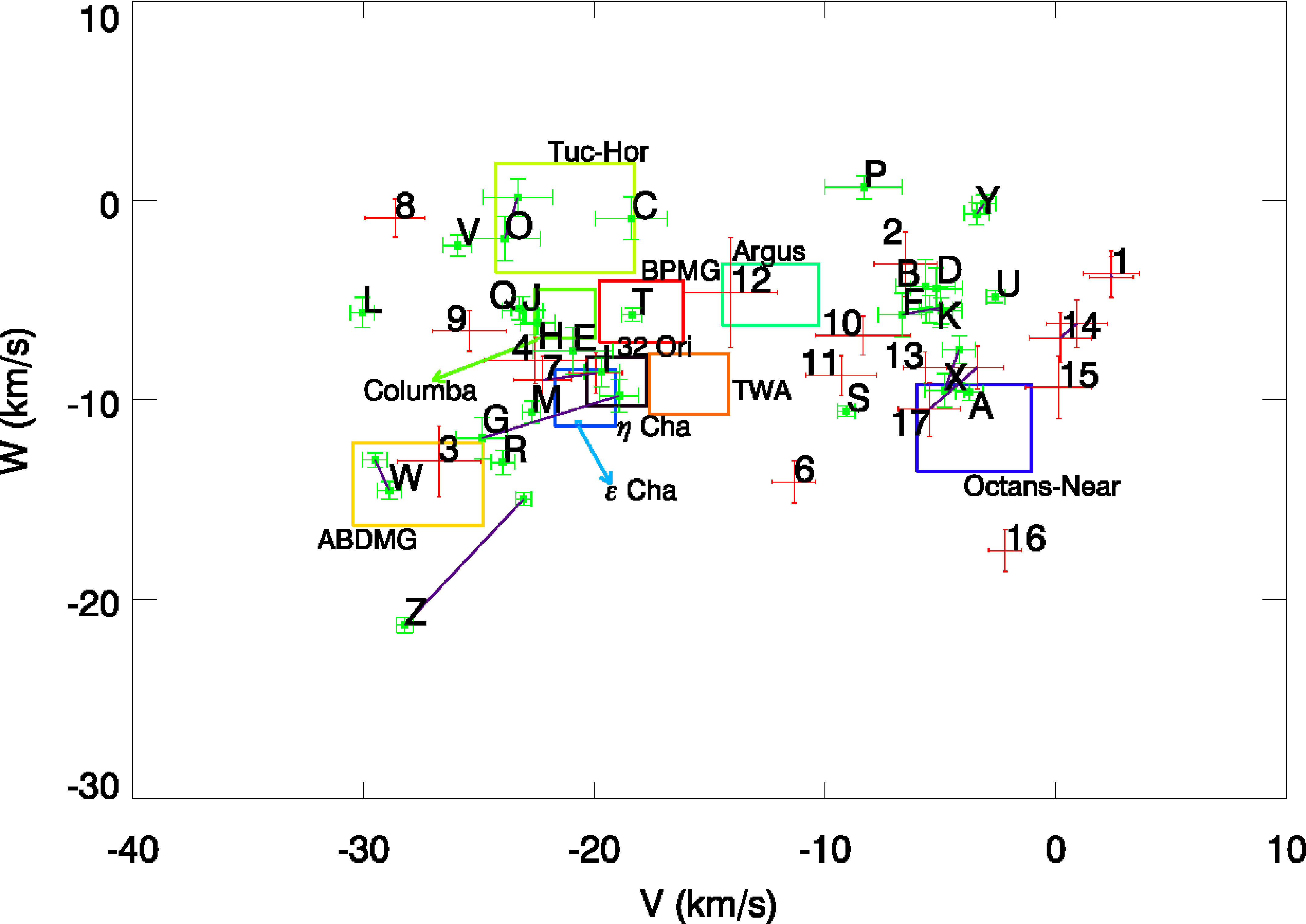}
    \end{center}
        \begin{flushleft}
 \caption{Boettlinger diagrams in $UVW$ highlighting the space motions of the characteristically young, likely-single objects in this paper (red symbols) and the likely-young sample from BJM15 (in green). Objects that lack a trigonometric parallax are connected by a solid purple line which indicate the error bar based on the photometric parallax range. Labels for this likely-young sample and the likely-young sample in BJM15 are the same as in Table~\ref{T_PX_LYS} and table~1 in BJM15, respectively.}
      \label{F_UVW}
    \end{flushleft}
    \end{figure*}

$UVW$ and $XYZ$ coordinates were calculated using the formulation in \cite{1987a_Johnson}, and are reported in columns 5, 6 and 7 of Table~\ref{T_Kin_LYS}. For objects without trigonometric parallaxes, a photometric parallax range is calculated by measuring the absolute magnitude corresponding to the extrema of the estimated age range of a target, using the \cite{2015a_Baraffe} models. Separate error bars are quoted for these: the first includes contributions from $\sigma_{\mu_{\alpha}}$, $\sigma_{\mu_{\delta}}$ and $\sigma_{\rm RV}$ and the second corresponds to half of the range in each velocity coordinate resulting from the extrema of the possible photometric parallaxes. The Boettlinger $U$ vs $V$ and $V$ vs $W$ diagrams are displayed in the upper and lower panels of Figure~\ref{F_UVW}, respectively, where for objects without trigonometric parallaxes a line connects $UVW$ points calculated at the extrema of the distances inferred from the photometric parallaxes. Overplotted on each Boettlinger diagram are boxes representing the 1$\sigma$ dispersion in $UVW$ for the 10 MGs introduced in BJM15 with the addition of the 32~Ori MG (\citealt{2007a_Mamajek,2017a_Bell}). Objects in red denote the 17 newly observed targets in this work and green objects, labelled A--Z, are the 26 likely-young objects from BJM15. We have improved the kinematic precision of our likely-young sample from BJM15 by incorporating TGAS trigonometric parallaxes for 15 objects.

{\scriptsize
\begin{table*}
\caption{Kinematic data for the likely-young sample in this work.}
\begin{center}
\begin{tabular}{lrrrrrr}
  \hline
  \hline
Target       &         $\mu_{\alpha}$ &         $\mu_{\delta}$ &                      $\pi$ &                  $U, X$ &                  $V, Y$ &                  $W, Z$ \\       
             & (${\rm mas\,yr}^{-1}$) & (${\rm mas\,yr}^{-1}$) &                      (mas) &              (\kms, pc) &              (\kms, pc) &              (\kms, pc) \\
\hline
1            &          $6.0 \pm 1.5$ &          $3.8 \pm 2.4$ &              $8.08, 11.29$ &  $-4.3 \pm 0.9 \pm 0.6$ &   $2.4 \pm 1.1 \pm 0.0$ &  $-3.8 \pm 1.1 \pm 0.1$ \\
             &                        &                        &                            &          $-17.6, -12.6$ &            $43.7, 61.0$ &         $-106.3, -78.0$ \\
2            &         $64.6 \pm 0.9$ &         $-5.6 \pm 0.9$ &           $20.80 \pm 0.38$ &         $-13.7 \pm 1.1$ &          $-6.5 \pm 1.4$ &          $-3.2 \pm 1.6$ \\
             &                        &                        &                            &         $-21.5 \pm 0.4$ &          $27.4 \pm 0.5$ &         $-33.1 \pm 0.6$ \\
3            &      $117.99 \pm 1.10$ &     $-127.99 \pm 0.44$ &           $26.60 \pm 0.32$ &          $-9.1 \pm 1.8$ &         $-26.7 \pm 1.8$ &         $-13.1 \pm 1.8$ \\
             &                        &                        &                            &         $-22.0 \pm 0.3$ &          $21.7 \pm 0.3$ &         $-21.5 \pm 0.3$ \\
4            &       $20.74 \pm 0.99$ &      $-27.41 \pm 1.15$ &           $13.48 \pm 0.24$ &         $-11.9 \pm 1.8$ &         $-22.6 \pm 2.2$ &          $-8.0 \pm 1.2$ \\
             &                        &                        &                            &       $-138.3 \pm 11.6$ &          $86.3 \pm 7.2$ &         $-16.9 \pm 1.4$ \\
5            &        $-42.5 \pm 1.2$ &         $-0.9 \pm 0.9$ &            $4.73 \pm 0.81$ &           $1.4 \pm 3.5$ &         $-28.1 \pm 4.8$ &          $38.9 \pm 5.0$ \\
             &                        &                        &                            &        $176.3 \pm 30.2$ &           $4.2 \pm 0.7$ &        $116.7 \pm 20.0$ \\
6            &        $2.56 \pm 0.65$ &        $4.38 \pm 1.02$ &            $2.90 \pm 0.24$ &         $-10.2 \pm 1.8$ &         $-11.3 \pm 0.9$ &         $-14.1 \pm 1.1$ \\
             &                        &                        &                            &          $65.4 \pm 5.4$ &        $275.4 \pm 22.8$ &        $196.9 \pm 16.3$ \\ 
7            &         $-7.4 \pm 2.0$ &        $-22.2 \pm 1.3$ &              $8.20, 10.38$ & $-10.5 \pm 1.5 \pm 0.8$ & $-21.1 \pm 1.2 \pm 1.2$ &  $-8.8 \pm 1.1 \pm 0.2$ \\
             &                        &                        &                            &            $76.3, 96.6$ &            $62.4, 76.3$ &            $32.2, 40.7$ \\
8            &        $-24.7 \pm 0.8$ &        $-12.2 \pm 1.6$ &            $8.30 \pm 0.24$ &         $-10.9 \pm 1.3$ &         $-28.6 \pm 1.3$ &          $-0.9 \pm 1.0$ \\
             &                        &                        &                            &          $75.1 \pm 2.2$ &          $80.6 \pm 2.3$ &          $48.7 \pm 1.4$ \\
9            &        $-16.2 \pm 0.8$ &         $11.3 \pm 0.5$ &           $10.29 \pm 0.22$ &          $-9.1 \pm 0.4$ &         $-25.4 \pm 1.6$ &          $-6.5 \pm 1.0$ \\
             &                        &                        &                            &          $14.9 \pm 0.3$ &          $82.3 \pm 1.8$ &          $49.6 \pm 1.1$ \\
10           &       $13.32 \pm 0.57$ &       $45.70 \pm 0.50$ &           $11.28 \pm 0.31$ &         $-29.7 \pm 1.5$ &          $-8.4 \pm 2.1$ &          $-6.8 \pm 1.0$ \\
             &                        &                        &                            &          $46.7 \pm 1.3$ &          $69.1 \pm 1.9$ &          $30.0 \pm 0.8$ \\
11           &         $15.6 \pm 0.9$ &        $-18.9 \pm 0.9$ &            $9.44 \pm 0.83$ &          $-1.6 \pm 1.2$ &          $-9.3 \pm 1.5$ &          $-8.8 \pm 1.0$ \\
             &                        &                        &                            &          $30.2 \pm 2.7$ &          $87.1 \pm 7.7$ &         $-52.2 \pm 4.6$ \\
12           &         $-0.2 \pm 0.9$ &         $-4.1 \pm 1.1$ &            $1.94 \pm 0.26$ &           $7.3 \pm 2.5$ &         $-14.1 \pm 2.0$ &          $-4.6 \pm 2.8$ \\
             &                        &                        &                            &         $-58.2 \pm 7.8$ &        $494.3 \pm 66.2$ &       $-134.2 \pm 18.0$ \\
13           &          $8.4 \pm 0.8$ &         $-6.1 \pm 0.6$ &            $4.50 \pm 0.35$ &          $-4.2 \pm 1.1$ &          $-5.7 \pm 0.9$ &          $-8.4 \pm 0.8$ \\
             &                        &                        &                            &         $-13.7 \pm 1.1$ &        $195.5 \pm 15.2$ &        $-104.7 \pm 8.1$ \\
14           &         $16.4 \pm 1.1$ &         $-0.6 \pm 1.4$ &              $8.02, 10.03$ &  $-7.7 \pm 0.6 \pm 0.8$ &   $0.6 \pm 1.3 \pm 0.4$ &  $-6.5 \pm 1.2 \pm 0.4$ \\
             &                        &                        &                            &            $-5.9, -4.8$ &            $75.3, 94.1$ &          $-81.5, -65.2$ \\
15           &         $19.3 \pm 0.4$ &          $1.7 \pm 1.3$ &            $7.20 \pm 0.73$ &         $-11.0 \pm 1.5$ &           $0.1 \pm 1.4$ &          $-9.4 \pm 1.6$ \\
             &                        &                        &                            &           $3.8 \pm 0.4$ &          $90.5 \pm 9.2$ &       $-105.3 \pm 10.7$ \\
16           &       $24.31 \pm 1.33$ &       $-8.53 \pm 0.58$ &            $8.82 \pm 0.41$ &          $-8.5 \pm 0.9$ &          $-2.2 \pm 0.7$ &         $-17.6 \pm 1.1$ \\
             &                        &                        &                            &           $6.3 \pm 0.3$ &          $54.1 \pm 2.5$ &         $-99.5 \pm 4.6$ \\
17           &         $16.3 \pm 2.3$ &         $-6.7 \pm 1.5$ &               $5.19, 6.85$ & $-10.2 \pm 1.7 \pm 1.3$ &  $-4.4 \pm 1.3 \pm 1.0$ &  $-9.4 \pm 1.2 \pm 1.0$ \\
             &                        &                        &                            &          $-51.2, -38.8$ &        $-159.4, -120.8$ &          $-95.1, -72.1$ \\
\hline
\label{T_Kin_LYS}
\end{tabular}
\end{center}
\end{table*}

\begin{table*}
		\contcaption{Kinematic data for the likely-young sample in BJM15, updated using new data from TGAS.}
\begin{center}
\begin{tabular}{lrrrrrr}
  \hline
  \hline
Target &         $\mu_{\alpha}$ &         $\mu_{\delta}$ &                      $\pi$ &                  $U, X$ &                  $V, Y$ &                  $W, Z$ \\       
       & (${\rm mas\,yr}^{-1}$) & (${\rm mas\,yr}^{-1}$) &                      (mas) &              (\kms, pc) &              (\kms, pc) &              (\kms, pc) \\
\hline
A      & $11.69 \pm 1.06$       & $-13.55 \pm 0.31$      & $8.33 \pm 0.30$            &          $-7.1 \pm 0.7$ &          $-3.8 \pm 0.6$ &          $-9.6 \pm 0.4$ \\
       &                        &                        &                            &         $-65.1 \pm 2.3$ &          $71.6 \pm 2.6$ &         $-71.1 \pm 2.6$ \\
B      & $13.29 \pm 1.25$       & $-4.90 \pm 0.37$       & $6.29 \pm 0.31$            &          $-9.4 \pm 1.4$ &          $-5.6 \pm 1.3$ &          $-4.3 \pm 1.3$ \\
       &                        &                        &                            &         $-87.6 \pm 4.3$ &          $83.5 \pm 4.1$ &        $-103.1 \pm 5.1$ \\
C      & $36.31 \pm 1.72$       & $-19.27 \pm 0.67$      & $8.96 \pm 0.44$            &         $-11.6 \pm 2.0$ &         $-18.4 \pm 1.6$ &          $-0.9 \pm 1.1$ \\
       &                        &                        &                            &         $-84.4 \pm 4.1$ &          $53.9 \pm 2.6$ &         $-49.3 \pm 2.4$ \\
D      & $10.31 \pm 0.80$       & $-7.06 \pm 0.34$       & $5.94 \pm 0.25$            &         $-12.2 \pm 2.0$ &          $-5.2 \pm 1.1$ &          $-4.4 \pm 1.1$ \\
       &                        &                        &                            &        $-137.9 \pm 5.8$ &          $66.3 \pm 2.8$ &         $-70.1 \pm 3.0$ \\
E      & $31.98 \pm 2.50$       & $-36.35 \pm 1.23$      & $8.82 \pm 0.60$            &         $-13.7 \pm 1.2$ &         $-20.9 \pm 1.7$ &          $-7.5 \pm 1.2$ \\
       &                        &                        &                            &         $-89.2 \pm 6.1$ &          $69.9 \pm 4.8$ &          $-1.5 \pm 0.1$ \\
F      & $12.4 \pm 1.6$         & $-14.6 \pm 1.7$        & $8.82-10.65$               &  $10.2 \pm 1.6 \pm 0.2$ &  $-5.8 \pm 1.0 \pm 0.8$ &  $-5.5 \pm 1.0 \pm 0.2$ \\
       &                        &                        &                            &          $-97.0, -80.4$ &            $31.4, 37.9$ &          $-44.8, -37.2$ \\
G      & $22.5 \pm 2.1$         & $46.2 \pm 2.1$         & $8.78-11.37$               &  $-6.6 \pm 0.6 \pm 0.2$ & $-21.9 \pm 1.0 \pm 3.0$ & $-10.9 \pm 0.9 \pm 1.1$ \\
       &                        &                        &                            &         $-100.4, -77.5$ &            $17.7, 23.0$ &          $-48.7, -37.6$ \\
H      & $27.2 \pm 1.3$         & $-52.5 \pm 1.3$        & $11.31 \pm 0.22$           &         $-13.4 \pm 0.6$ &         $-22.5 \pm 0.7$ &          $-6.1 \pm 0.6$ \\
       &                        &                        &                            &         $-87.2 \pm 1.7$ &          $11.0 \pm 0.2$ &         $-10.0 \pm 0.2$ \\
I      & $10.0 \pm 1.2$         & $-48.0 \pm 1.2$        & $11.24 \pm 0.22$           &         $-12.1 \pm 0.5$ &         $-19.7 \pm 0.8$ &          $-8.6 \pm 0.7$ \\
       &                        &                        &                            &         $-88.3 \pm 1.7$ &          $-1.9 \pm 0.1$ &         $-11.1 \pm 0.2$ \\
J      & $-21.03 \pm 0.90$      & $-42.80 \pm 0.78$      & $8.87 \pm 0.31$            &         $-13.7 \pm 0.6$ &         $-23.1 \pm 0.9$ &          $-5.5 \pm 0.7$ \\
       &                        &                        &                            &         $-87.8 \pm 3.1$ &         $-13.9 \pm 0.5$ &          $69.3 \pm 2.4$ \\
K      & $-6.9 \pm 1.6$         & $7.4 \pm 1.6$          & $7.06-8.39$                & $-10.7 \pm 1.0 \pm 0.9$ &  $-5.2 \pm 1.0 \pm 0.3$ &  $-5.5 \pm 0.7 \pm 0.0$ \\
       &                        &                        &                            &            $-7.1, -6.0$ &            $40.0, 47.6$ &          $112.0, 133.2$ \\
L      & $-116.16 \pm 0.69$     & $-68.95 \pm 0.86$      & $18.59 \pm 0.45$           &         $-16.4 \pm 0.5$ &         $-30.1 \pm 0.5$ &          $-5.6 \pm 0.8$ \\
       &                        &                        &                            &          $27.7 \pm 0.7$ &         $-15.8 \pm 0.4$ &          $43.4 \pm 1.0$ \\
M      & $-37.07 \pm 0.54$      & $-19.51 \pm 0.75$      & $10.84 \pm 0.27$           &          $-8.7 \pm 0.5$ &         $-22.7 \pm 0.4$ &         $-10.6 \pm 0.6$ \\
       &                        &                        &                            &          $22.1 \pm 0.6$ &          $30.1 \pm 0.8$ &          $84.3 \pm 2.1$ \\
N      & $39.4 \pm 1.6$         & $-37.7 \pm 1.6$        & $25.25-33.18$              & $-25.0 \pm 0.4 \pm 1.0$ & $-37.2 \pm 0.4 \pm 0.1$ & $-52.6 \pm 0.5 \pm 0.7$ \\
       &                        &                        &                            &            $13.6, 18.0$ &            $16.1, 21.2$ &            $19.9, 26.3$ \\
O      & $16.6 \pm 1.7$         & $-7.5 \pm 1.7$         & $7.45-9.42$                &  $-8.3 \pm 1.0 \pm 0.5$ & $-23.7 \pm 1.5 \pm 0.3$ &  $-1.3 \pm 1.1 \pm 1.0$ \\
       &                        &                        &                            &            $16.9, 21.5$ &           $93.3, 118.7$ &          $-43.2, -33.9$ \\
P      & $-135.0 \pm 1.1$       & $55.2 \pm 1.1$         & $21.71 \pm 1.64^{\rm a}$   &         $-31.2 \pm 1.8$ &          $-8.3 \pm 1.7$ &           $0.7 \pm 0.6$ \\
       &                        &                        &                            &           $8.2 \pm 0.6$ &           $1.3 \pm 0.1$ &          $45.3 \pm 3.4$ \\
Q      & $-63.12 \pm 0.77$      & $-72.02 \pm 1.41$      & $20.18 \pm 0.55$           &          $-7.0 \pm 0.4$ &         $-23.3 \pm 0.3$ &          $-5.4 \pm 0.6$ \\
       &                        &                        &                            &          $38.9 \pm 1.1$ &           $3.9 \pm 0.1$ &          $30.4 \pm 0.8$ \\
R      & $-11.87 \pm 0.50$      & $-16.25 \pm 0.78$      & $7.66 \pm 0.27$            &          $-6.5 \pm 0.7$ &         $-24.0 \pm 0.5$ &         $-13.1 \pm 0.6$ \\
       &                        &                        &                            &          $65.4 \pm 2.3$ &          $71.5 \pm 2.5$ &          $87.5 \pm 3.1$ \\
S      & $-3.33 \pm 0.83$       & $108.85 \pm 1.01$      & $20.44 \pm 0.38$           &         $-40.7 \pm 0.5$ &          $-9.1 \pm 0.4$ &         $-10.6 \pm 0.3$ \\
       &                        &                        &                            &          $28.5 \pm 0.5$ &          $30.9 \pm 0.6$ &          $25.1 \pm 0.5$ \\
T      & $-35.58 \pm 0.65$      & $55.37 \pm 0.64$       & $22.56 \pm 0.24$           &         $-21.5 \pm 0.3$ &         $-18.4 \pm 0.4$ &          $-5.7 \pm 0.3$ \\
       &                        &                        &                            &          $17.9 \pm 0.2$ &         $ 32.6 \pm 0.3$ &          $24.1 \pm 0.3$ \\
U      & $3.51 \pm 0.54$        & $59.64 \pm 0.66$       & $14.57 \pm 0.23$           &         $-26.1 \pm 0.4$ &          $-2.6 \pm 0.4$ &          $-4.8 \pm 0.3$ \\
       &                        &                        &                            &          $37.1 \pm 0.6$ &          $47.2 \pm 0.7$ &          $33.3 \pm 0.5$ \\
V      & $-17.88 \pm 0.69$      & $3.96 \pm 0.91$        & $8.60 \pm 0.29$            &         $-10.9 \pm 0.5$ &         $-25.9 \pm 0.6$ &          $-2.2 \pm 0.5$ \\
       &                        &                        &                            &          $42.4 \pm 1.4$ &          $96.0 \pm 3.2$ &          $50.0 \pm 1.7$ \\
W      & $30.9 \pm 1.5$         & $7.1 \pm 1.5$          & $16.61-21.26$              & $-12.1 \pm 0.4 \pm 0.6$ & $-29.2 \pm 0.5 \pm 0.3$ & $-13.8 \pm 0.4 \pm 0.8$ \\
       &                        &                        &                            &            $10.7, 13.7$ &            $44.5, 57.0$ &            $10.7, 13.7$ \\
X      & $18.9 \pm 1.5$         & $-7.0 \pm 1.5$         & $7.31-8.98$                &  $-7.5 \pm 0.7 \pm 0.6$ &  $-4.5 \pm 0.8 \pm 0.3$ &  $-8.5 \pm 0.8 \pm 1.0$ \\
       &                        &                        &                            &            $72.6, 89.2$ &             $5.9, 73.7$ &          $-73.2, -59.5$ \\
Y      & $38.1 \pm 1.3$         & $16.2 \pm 1.3$         & $11.31-14.84$              & $-15.1 \pm 0.5 \pm 2.0$ &  $-3.3 \pm 0.5 \pm 0.2$ &  $-0.4 \pm 0.5 \pm 0.2$ \\
       &                        &                        &                            &              $0.1, 0.2$ &            $55.8, 72.9$ &          $-50.1, -38.3$ \\
Z      & $105.7 \pm 2.0$        & $-85.7 \pm 1.4$        & $17.58-23.06$              & $-11.4 \pm 0.5 \pm 1.5$ & $-25.6 \pm 0.3 \pm 2.6$ & $-18.1 \pm 0.4 \pm 3.2$ \\
       &                        &                        &                            &            $-0.3, -0.2$ &            $33.7, 44.2$ &          $-35.9, -27.3$ \\
\hline
\end{tabular}
\begin{flushleft}
a) Parallax from \protect\cite{2012a_de_Bruijne}.
\end{flushleft}
\end{center}
\end{table*}}

{\tiny
\begin{table*}
  \caption{Kinematic properties of the Pisces MG. Column 2 is the number of proposed members. Columns 7 and 8 correspond to the convergent points in right ascension and declination, respectively.}
\begin{center}
\begin{tabular}{lrrrrrrr}
  \hline
  \hline
Name     & $N$ & Age       & Distance & $UVW$                 & $XYZ$                 & $\alpha_{\rm CP}$   & $\delta_{\rm CP}$ \\
         &     &           &          & $\sigma_{UVW}$        & $\sigma_{XYZ}$        &                     &                   \\
         &     & (Myr)     & (pc)     & (\kms)                & (pc)                  & ($^{\circ}$)        & ($^{\circ}$)      \\
  \hline
Pisces   &  14 &  $30-50$? & $48-222$ &    $-8.8, -3.9, -6.3$ &  $-41.7, 88.8, -77.1$ &            $69.344$ &          $-7.634$ \\
         &     &           &          &       $3.9, 3.2, 3.0$ &    $58.2, 51.6, 28.5$ &                     &                   \\
\hline
\end{tabular}
\end{center}
\label{T_Pisces_Kinematics}
\end{table*}}

\section{A new co-eval moving group?}\label{S_Pisces}

In BJM15 we identified 7 stars (A, B, D, F, K, X and Y) with ages between 30 and 200\,Myr and with dispersions $< 5\,{\rm km\,s}^{-1}$ in each Galactic space velocity component. The $UVW$ are similar to the Octans-Near MG reported by \cite{2013a_Zuckerman}, but with velocities $\sim 5\,$\kms~more positive in $U$ and $\sim 10\,$\kms~more negative in $W$, on average (see Figure~\ref{F_UVW}). Since 5 of these have sky positions in or in the vicinity of Pisces, we dubbed this potential new group as the ``the Pisces moving group''.

In this work, we identify some potential new members of the Pisces MG and re-evaluate the membership status of the previously reported seven objects in light of new astrometric data from TGAS. There appear to be seven new objects that have broadly similar kinematics to the seven objects reported in BJM15. These are targets 1, 2, 11, 13, 14, 15 and 17. If we treat all fourteen objects as part of the same group, which we call the ``Pisces sample'' then the measured $UVW$ velocity of the group would be: $U = -8.8 \pm 3.9 (\pm 2.9); V = -3.9 \pm 3.2 (\pm 1.1); W = -6.3 \pm 3.0 (\pm 1.1)$\,\kms. With the exception of target 17, none are within the 1\,$\sigma$ dispersion in all 3 $UVW$ coordinates for Octans-Near.

\begin{figure*}
    \begin{center}
            \includegraphics[width=0.7\textwidth]{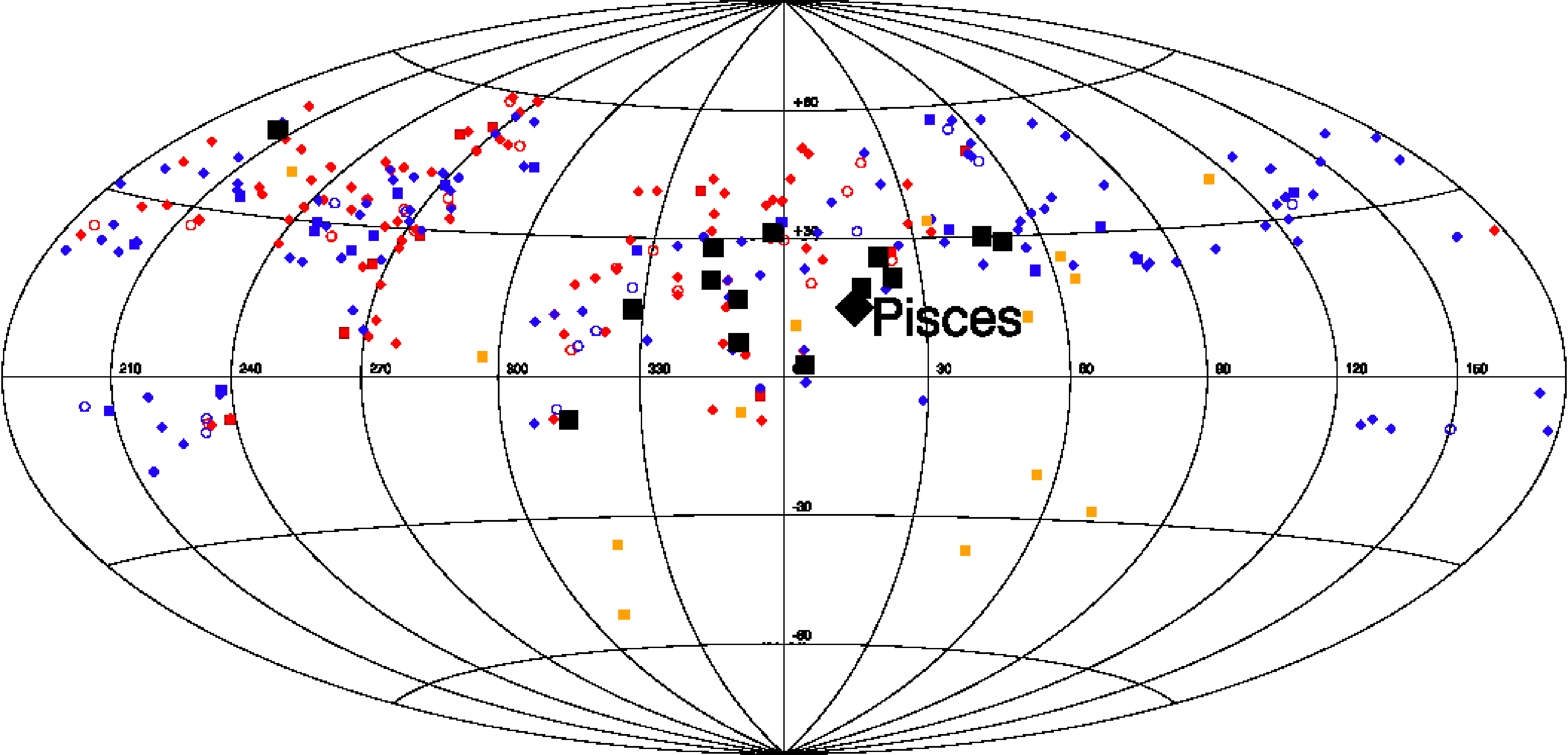}
    \end{center}
        \begin{flushleft}
 \caption{An Aitoff map projecting $\alpha$ and $\delta$ for Pisces objects (black squares), Octans-Near objects (gold squares, from \citealt{2013a_Zuckerman}), other likely-young objects in this work (red squares), and in BJM15 (blue squares), likely TLSPB systems from this work (red open circles) and in BJM15 (blue open circles), and all other objects from this work (red diamonds) and in BJM15 (blue diamonds). The large black diamond is the central point of the Pisces constellation.}
      \label{F_Aitoff}
    \end{flushleft}
\end{figure*}

Only seven of the Pisces sample have a trigonometric parallax, which are used to examine the spatial distribution of the group. The $XYZ$ positions are calculated as: $X = -41.7 \pm 58.2 (\pm 2.4); Y = 88.8 \pm 51.6 (\pm 6.0); Z = -77.1 \pm 28.5 (\pm 5.0)$\,pc. The mean $X$ position of the Pisces sample is similar to that of the Octans-Near MG, however both the $Y$ and $Z$ position are larger by $\sim 1\,\sigma$. Whilst the $V$ velocity is in excellent agreement with Octans-Near, the $U$ and $W$ velocities of our Pisces sample are both $\sim 1\,\sigma$ smaller and we do not suggest the two groups are either spatially or kinematically coherent. Some doubt may be cast on the commonality of these objects since their distances range from 48.0 (target 2) to 222.2\,pc (target 13), but we note the other 5 objects with parallaxes are much more compact -- between 105.9 and 168.3\,pc. With the exception of target 2, all objects with a trigonometric parallax in the Pisces sample lie beyond the most distant Octans-Near object (which are between 24 and 94\,pc). We note that given our measured mean $Z$ position for the Pisces sample, this would represent, in Galactic coordinates, the most southerly MG. Excluding target K (RA = 13h 37m 09s, DEC = 44d 44m 55s) the sky positions of the Pisces sample are distributed around the $\alpha$ and $\delta$ point (23h 19m 12s, 08d 30m 00s) with a dispersion of 15 and 13 degrees, respectively (see Figure~\ref{F_Aitoff}). 

\begin{figure*}
    \begin{center}
            \centering
            \includegraphics[width=1.0\textwidth]{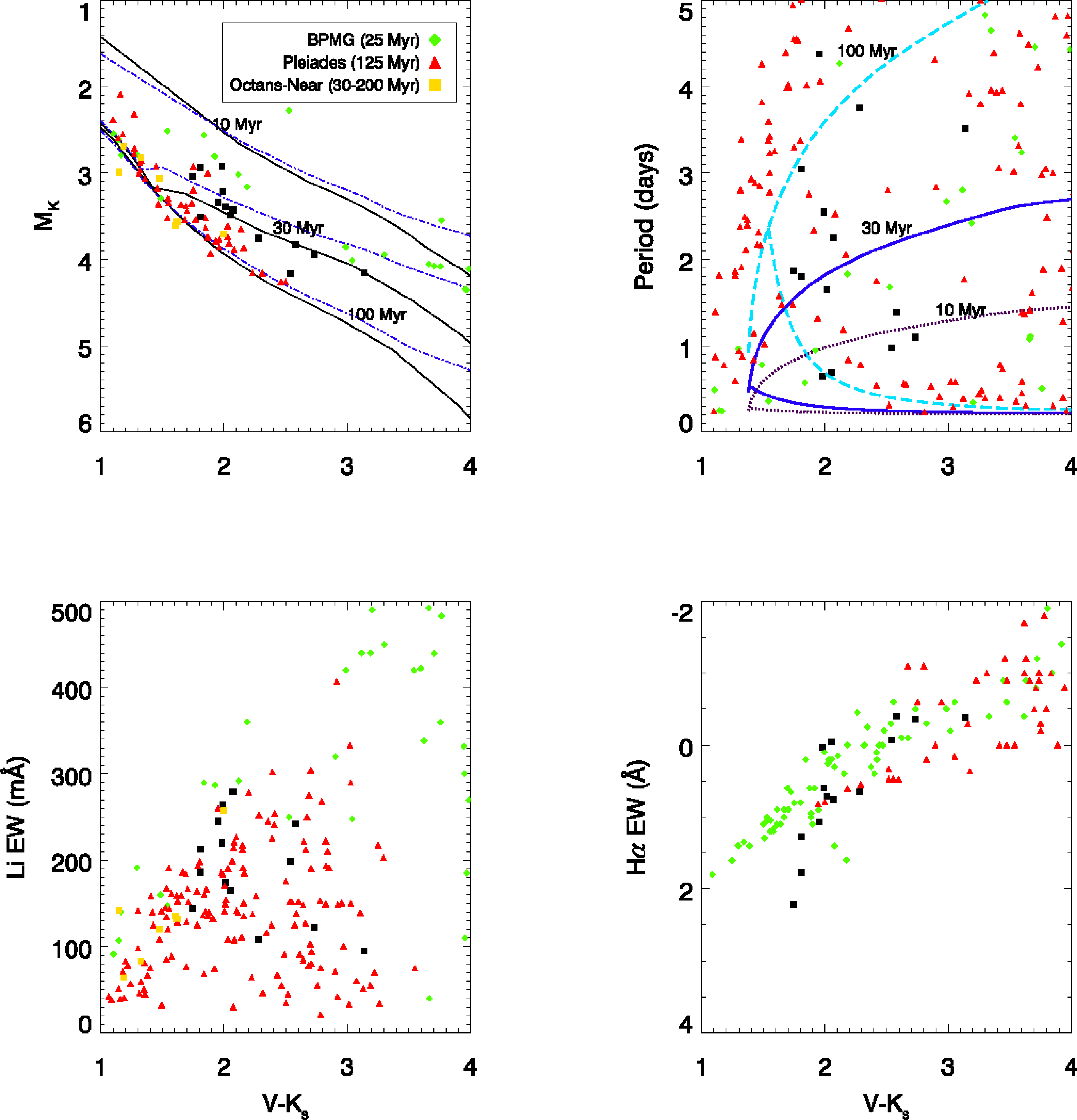}
    \end{center}
        \begin{flushleft}
 \caption{Distributions for the 14 objects that form the Pisces group (black squares) as a function of \VK~in terms of top-left: the CMD, where blue dotted and solid black lines represent isochrones from \protect\cite{2015a_Bell} and \protect\cite{2015a_Baraffe}, respectively; top-right: rotation-periods, where the various lines denote the same gyrochrones as in Figure~\ref{F_Gyrochrones}; bottom-left: Li EW and bottom-right: H$\alpha$ EW. Green diamonds represent objects in the BPMG from \protect\cite{2016a_Messina}, except for the bottom-right panel, where H$\alpha$ EWs are from objects in the IC~2391, NGC~2602 (30 and 50\,Myr, respectively, \protect\citealt{2001a_Randich}) and NGC~2547 (35\,Myr, \protect\citealt{2000a_Jeffries, 2005a_Jeffries}), red triangles are Pleiades members where \MK~and rotation periods are from table~2 in \protect\cite{2016a_Rebull}, Li EW data are the same as in Figure~\ref{F_Li_EW} and H$\alpha$ EW data are from \protect\cite{1995a_Hodgkin} and gold squares are Octans-Near members from \protect\cite{2013a_Zuckerman}.}
      \label{F_Pisces}
    \end{flushleft}
\end{figure*}

Figure~\ref{F_Pisces} displays (as a function of \VK) the distribution in \MK, rotation period, Li and H$\alpha$~EW for the possible Pisces MG members and compares them with members of the BPMG and the Pleiades. Also overplotted as blue open squares (in the \MK~and Li plots) are the sample of Octans-Near members, where the data are sourced directly from \cite{2013a_Zuckerman}. The CMD suggests the Pisces sample may be closer in age to the BPMG and appears to be younger than Octans-Near, however the Li EW distribution is better matched to the Pleiades sample. Neither gyrochronology or H$\alpha$ can be used to resolve ages between 20 and 125\,Myr. At present we cannot find a more precise method to calculate individual ages than the Li EW technique. There are 6 G-type and 8 K-type stars in the sample ranging from G1 (target B) to K4 (targets 1 and Y).

To test for co-evality in the Pisces sample we fit isochronal models from 5\,Myr to 200\,Myr (with a 1\,Myr resolution) from \cite{2015a_Baraffe} to the \MK~versus~\VK~CMD for our 14 Pisces candidates, to members of BPMG, to Pleiads, to Octans-Near members, to the remaining 29 objects in the likely-young sample, and finally to a combined sample of Pisces and Octans-Near. In each case the best fit isochrone is defined as that which minimises the median absolute deviation (MAD) in~\MK. This approach is robust to non-members and is suitable given that the mean error in \MK~amongst targets in all 4 samples are similar ($\sigma_{M_{\rm K_{\rm s}}} = 0.10 \pm 0.02$). For Pisces this minimum occurs at $\sim 40$\,Myr, where the MAD is $\sim 0.1$\,mag. This dispersion is smaller than the MAD amongst BPMG members (MAD $\sim$~0.25\,mag at 20\,Myr), similar to the Pleiades (distance = $134 \pm 6$ (TGAS), MAD $\sim 0.1$\,mag, $> 70$\,Myr), and much less than the dispersion amongst Octans-Near members (MAD $\sim 0.6$\,mag, 50--100\,Myr), whereas the rest of the likely-young sample have a minimum MAD of $\sim 0.25$\,mag at $\sim 50$\,Myr. The Pisces/Octans-Near sample have a minimum MAD of $\sim 0.8$\,mag which strongly suggests the two samples are not co-eval. Since the dispersion amongst Pisces candidates about a single isochrone is similar to that observed in the co-eval Pleiades and is less than the scatter observed in a population of non-kinematically related young stars, and less than the scatter found for BPMG, which is usually assumed to be a co-eval group, these results suggest Pisces could be a co-eval group with an age of $30-50$\,Myr. A similar approach using Li abundances was inconclusive because of the large scatter in Li amongst all populations.

Figure~\ref{F_UVW} appears to show 2 broad populations in both Boettlinger diagrams. One region which contains $\epsilon$ and $\eta$~Cha, BPMG, 32~Ori, Columba, Tuc-Hor, Carina and the ABDMG, i.e. the ``local association'', which are largely encapsulated within the $UVW$ ``good-box'' ( $-15 < U < 0, -34 < V < -10, -20 < W < +3$\,\kms) defined by \cite{2004a_Zuckerman}. The second population, that includes Octans-Near and Pisces, appears to be similar in $U$ and $W$ space ($-10$ and $-8$\,\kms, respectively), but with a $V$ velocity $\sim 15-25\,$\kms~more positive.

We test for a common origin of the candidate members of the Pisces MG by taking their present $UVWXYZ$ and tracing back their Galactic orbits under the \cite{1991a_Allen} Galactic potential (source code provided by B. Pichardo) in timesteps of $-0.1\,$Myr. The experiment is performed for three separate populations: a) the 7 Pisces MG candidates with a trigonometric parallax, b) the entire sample of 14 Pisces MG candidates and c) the 14 Pisces MG candidates with the 14 Octans-Near members reported in \cite{2013a_Zuckerman}. For the Octans-Near stars we use the $UVW$ and errors provided in \cite{2013a_Zuckerman} and choose trigonometric parallaxes from either Gaia or the reduced Hipparcos data (\citealt{2007a_van_Leeuwen}). Uncertainties are simulated by generating 100 simulations for each star perturbed from their mean $UVWXYZ$ from a normal distribution in their initial errors and calculating the dispersion in $XYZ$ for each timestep.

\begin{figure*}
    \begin{center}
            \includegraphics[width=0.9\textwidth]{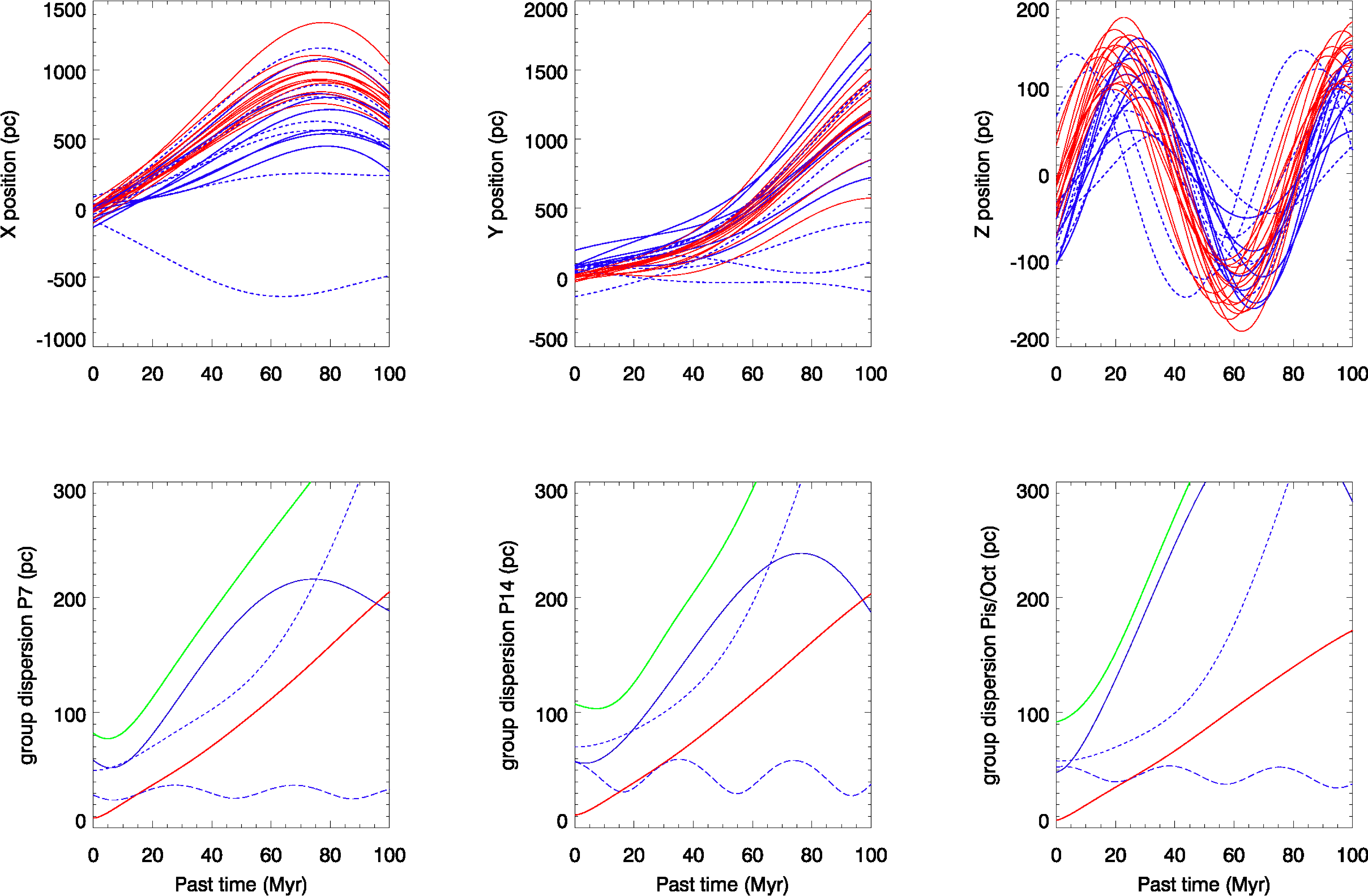}
    \end{center}
     \begin{flushleft}
\caption{Top: Kinematic tracebacks for a) Pisces candidates with trigonometric parallaxes, b) All Pisces candidates and c) All Pisces candidates and Octans-Near members from \protect\cite{2013a_Zuckerman}, in left: $X$, middle: $Y$, and right: $Z$ position. Bottom: total dispersion in $X, Y, Z$ positions (given by blue solid, dotted and dashed lines, respectively), quadrature sum of the dispersions ($\sqrt{X^{2} + Y^{2} + Z^{2}}$, green solid lines) and the error in the measurement amongst the stars in the sample (red solid lines) as a function of past time amongst groups a (left), b (middle) and c (right). For experiments a and b (Pisces candidates) between 30 and 50\,Myr the measurement error is about half of the total distance dispersion.}
 \label{F_TB}
     \end{flushleft}
\end{figure*}

Figure~\ref{F_TB} displays the results of these 3 experiments. The left-panels show the tracebacks for each star, using only their mean $UVWXYZ$ inputs. Blue solid, blue dashed and red solid lines represent the objects in experiments a, b and c, respectively. There does not appear to be any time in the past where these stars occupied a smaller volume. The clearly anomalous star in the X-position diagram is star F, which lacks a trigonometric parallax. The limitations of these tracebacks are evident from the lower panels, where dispersions in X (blue solid), Y (blue dot), Z (blue dash), and total volume ($=\sqrt{X^2 + Y^2 + Z^2}$, green solid) are plotted as a function of past-time. The solid red line represents the median of the volume dispersions of the 100 traceback iterations for each star. The volume error increases with time, and the expected dispersion just caused by uncertainties in the starting measurements for the stars is $> 70\,$pc by $\sim 40\,$Myr in each experiment, the presumed age of the Pisces MG, therefore at present one cannot perform a reliable traceback on these objects.

We suggest that we may be uncovering a similar kinematic entity to the LA, but with a different set of Galactic space velocities. Since these are all younger than 200\,Myr and have no kinematic similarity with objects in the LA we speculate this is another young population of stars within 200\,pc that have broadly similar kinematics, and are possibly composed of sub-structures similar to the MGs discovered in the LA. Pisces and Octans-Near in a broad sense are as similar to each other as BPMG and ABDMG in terms of ages and kinematics. The objects we find in this work lie on the boundaries of $UVW$ that determine the young disc population as defined in Eggen (1984; 1989).

\nocite{1984a_Eggen}
\nocite{1989a_Eggen}

\section{Conclusions}\label{S_Conclusion}

Continuing from our work in BJM15, we spectroscopically observed 122 rapidly rotating, X-ray active FGK stars to measure their kinematics and estimate their ages. In this work we initially observed 19 targets with positions and proper-motions suggesting they might be further members of the tentative Pisces MG propsed by BJM15, together with a further 103 targets that were chosen without kinematic bias. We measure radial velocities (RVs) to precisions better than 5\kms~for 109 stars, of which 19 are probably in tidally-locked short-period binary systems (TLSPBs) identified from large RV variations in 2 or more measurements. Only 41 targets had 2 or more RV measurements, but many of the other targets with only one RV measurement and no spectroscopic signs of youth are likely to be TLSPBs, and this is supported by the positions of a subsample of them in the absolute magnitude vs colour diagram. We use Li as the primary age indicator for our stars and additional age indicators support the Li age for targets younger than 200\,Myr. In total we identified 17 likely-single stars younger than 200\,Myr.

Combining these with the likely-young sample in BJM15, and updating their kinematics using new astrometric data from the Tycho-Gaia Astrometric Solution catalogue, we have 43 likely-single stars younger than 200\,Myr, of which 39 are probably between 30 and 200\,Myr, 3 could be as young as 25\,Myr and only one star is probably younger than 20\,Myr. Comparisons with solar-type stars in known nearby MGs suggests there is a possible overdensity of very young stars ($< 25$\,Myr) within 100\,pc of the Sun compared to the volumes sampled by our survey, which would certainly have been capable of finding such stars. This will be investigated in more detail in a forthcoming paper.

We have found a kinematically coherent group of 14 stars that appear distinct from most MGs. The absolute colour-magnitude diagram and Li depletion suggests co-evality and an age of 30--50\,Myr. Most of the stars are towards, or close to, the constellation of Pisces and we dub this the Pisces moving group. The Pisces MG shares similar, but not identical, UVW to the Octans-Near group; however, members of the latter group are much nearer to the Sun and older than members of the Pisces MG.

Several hundreds of potentially young targets remain in our catalogue and are yet to be observed spectroscopically. We aim to incorporate previously observed objects and future observations with forthcoming $\mu$as precision Gaia DR2 data (expected in 2018) to further investigate the demography of the young stars in the Solar neighbourhood.

\section{Acknowledgements}\label{S_Acknowledgements}

The authors thank the anonymous referee, whose comments and suggestions improved the quality of this manuscript. ASB acknowledges the financial support from the Consejo Nacional de Ciencia y Tecnolog\'ia (M\'exico) and the United Kingdom Science and Technology Funding Council (STFC). RDJ acknowledges financial support from the STFC. JLW acknowledges financial support from STFC and Sonderforschungsbereich SFB 881 "The Milky Way System" (subproject P1) of the German Research Foundation (DFG). Based on observations made with the Isaac Newton Telescope operated on the island of La Palma by the Isaac Newton Group of Telescopes in the Spanish Observatorio del Roque de los Muchachos of the Instituto de Astrof\'isica de Canarias. This research has made use of the NASA/IPAC Infrared Science Archive, which is operated by the Jet Propulsion Laboratory, California Institute of Technology, under contract with the National Aeronautics and Space Administration. This work is based [in part] on observations made with the Spitzer Space Telescope, which is operated by the Jet Propulsion Laboratory, California Institute of Technology under a contract with NASA. This publication makes use of data products from the Two Micron All Sky Survey, which is a joint project of the University of Massachusetts and the Infrared Processing and Analysis Center/California Institute of Technology, funded by the National Aeronautics and Space Administration and the National Science Foundation. This research is based on observations with AKARI, a JAXA project with the participation of ESA. This research has made use of the SIMBAD database, operated at CDS, Strasbourg, France.

\appendix
\section{Tidally-locked short-period binaries}\label{S_TLSPBs}

In $\S$\ref{SS_Binary} we describe our binary scoring system based on RV measurements. Here we display the basic properties of the 33 objects that we score 5, and are hence likely to be part of TLSPBs. Nineteen are from BJM15 and 14 are from this work. We measure 22 of these to have Li abundances consistent with ages younger than 300\,Myr. The fraction of young stars in the TLSPB sample is much larger than the whole spectroscopically observed sample because during observations their spectra were noted to have significant Li absorption and were reobserved on a later night and found to have inconsistent RVs. Generally, targets that were not found to exhibit Li did not have a second spectrum taken, but could still be TLSPBs following subsequent RV measurements.

{\scriptsize
\begin{table*}
  \caption{Basic properties of the 33 targets likely to be TLSPBs. Ages are estimated from Li EWs. RVs taken from: a) \protect\cite{2012a_de_Bruijne}, b) \protect\cite{2013a_Kordopatis}, c) \protect\cite{2009a_Mermilliod}, d) \protect\cite{2006a_Gontcharov}, e) \protect\cite{2014a_Malo}.}
\begin{center}
\begin{tabular}{p{2.5cm}p{0.9cm}p{0.9cm}p{0.8cm}p{1.8cm}p{1.8cm}p{1.7cm}p{0.8cm}p{1.1cm}p{1.4cm}}
\hline
\hline
SuperWASP ID/Label   & $V$    & $V-K_{\rm s}$ & Period & RV$_{1}$         & RV$_{2}$         & $\pi$            & H$\alpha$ EW & Li EW        & Age      \\
(1SWASP J-)          & (mag)  & (mag)         & (days) & (\kms)           & (\kms)           & (mas)            & (\AA)        & (m\AA)       & (Myr)     \\
\hline
\multicolumn{10}{c}{TLSPBs from BJM15} \\
010822.21$+$313801.1 & 11.675 &         1.924 &  0.730 & $-10.2 \pm 3.1$  & $+9.1 \pm 3.3$   & $5.15-6.10$      &     $+0.47$ & $213 \pm 27$ &  $30-150$ \\	
033028.34$+$541737.6 &  9.190 &         1.910 &  0.790 & $+66.4 \pm 0.6$  & $+33.93^{\rm a}$ & $10.65 \pm 0.28$ &     $+2.09$ & $38 \pm 18$  & $500-700$ \\
033850.51$+$463612.6 & 10.069 &         1.312 &  0.711 & $+2.5 \pm 3.6$   & $-7.5 \pm 3.9$   & $5.18 \pm 0.29$  &     $+2.58$ & $< 27$       & $> 1000$  \\
083336.39$+$322444.7 & 11.648 &         2.444 &  2.505 & $+39.1 \pm 6.4$  & $+77.1 \pm 2.5$  & $2.94 \pm 0.37$  &     $+0.43$ & $104 \pm 18$ & $100-200$ \\
                     &        &               &        & $+45.3 \pm 2.9$  &                  &                  &              &              &           \\
100034.85$-$085448.3 & 11.502 &         2.428 &  4.218 & $+57.2 \pm 2.7$  & $+75.6 \pm 3.5$  & $10.47$          &     $-0.15$ & $< 5$        & $> 1000$  \\
132943.20$-$045422.1 &  9.783 &         2.132 &  3.025 & $+31.8 \pm 1.8$  & $-75.2 \pm 0.3$  & $5.42 \pm 0.78$  &     $+0.69$ & $< 10$       & $> 1000$  \\
152808.32$-$101035.0 & 11.460 & 2.747 & 1.430 & $-40.2 \pm 11.2$ & $+18.8^{\rm b}, -27.8^{\rm b}$ & $7.70 \pm 0.27$ & $+0.21$ & $< 7$        & $> 1000$  \\
                     &        &               &        & $+35.6^{\rm b}, +23.2^{\rm b}$    & &                  &              &              &           \\
153144.23$-$073454.3 & 10.774 &         1.883 &  1.377 & $-58.0 \pm 0.9$                   & $-24.5^{\rm b}, -5.9^{\rm b}$ & $6.87 \pm 0.23$  &     $+0.87$ & $< 8$        & $> 1000$  \\
162641.33$+$335041.8 &  9.622 &         2.492 & 19.880 & $-119.5 \pm 0.4$ & $-125.3 \pm 0.3$ & $2.08 \pm 0.22$  &     $+0.55$ & $31 \pm 3$   & $100-300$ \\
174705.04$+$332129.1 & 11.456 &         2.495 &  3.169 & $+10.6 \pm 0.4$  & $-14.4 \pm 0.3$  & $8.90 \pm 0.25$  &     $+0.05$ & $208 \pm 36$ & $30-150$  \\
204859.58$-$064453.6 &  9.588 &         2.318 &  2.049 & $-27.9 \pm 2.1$  & $-50.6 \pm 0.9$  & $8.19 \pm 0.40$  &     $+0.09$ & $50 \pm 1$   & $100-300$ \\
210707.11$+$063232.1 &  9.869 &         2.000 &  6.331 & $+4.6 \pm 0.4$   & $-0.9 \pm 0.4$   & $3.93 \pm 0.48$  &     $+1.15$ & $57 \pm 8$   & $> 1000$  \\
212135.86$+$094835.3 & 10.333 &         1.904 &  3.564 & $+12.0 \pm 0.6$  & $-54.0 \pm 0.7$  & $3.06 \pm 0.52$  &     $+0.73$ & $32 \pm 4$   & $200-700$ \\
214809.40$+$191012.9 & 10.734 &         2.967 &  1.165 & $-86.0 \pm 0.5$  & $-71.8 \pm 0.7$  & $25.62$          &     $-0.20$ & $21 \pm 2$   & $100-300$ \\
\hline
\multicolumn{10}{c}{TLSPBs from this work} \\
002334.66$+$201428.6 & 10.842 &         3.505 &  1.057 & $+0.2 \pm 2.1$   & $-3.4 \pm 1.3$   & $15.36-26.57$    &     $-0.61$ & $325 \pm 40$ & $5-30$    \\
010448.62$+$404051.0 & 11.344 &         2.008 &  1.628 & $+23.6 \pm 2.6$  & $-20.5 \pm 1.2$  & $7.64 \pm 0.28$  &     $+0.39$ & $200 \pm 29$ & $30-200$  \\
012557.65$+$471324.5 & 10.854 &         2.412 &  1.519 & $+32.3 \pm 2.6$  & $-2.2 \pm 1.2$   & $4.08 \pm 0.41$  &     $-2.02$ & $54 \pm 28$  & $100-300$ \\
013626.71$+$250828.4 & 10.846 &         3.387 &  3.938 & $+13.5 \pm 0.7$  & $-15.2 \pm 0.9$  & $5.54 \pm 0.29$  &     $-3.38$ & $81 \pm 21$  & $30-200$  \\
122940.92$+$243114.6 & 13.648 &         2.923 &  3.670 & $-38.9 \pm 1.2$  & $+0.30 \pm 0.12^{\rm c}$ & $11.42 \pm 0.30$ &     $+0.64$ & $35 \pm 6$   & $> 1000$  \\
141630.88$+$265525.1 & 10.489 &         2.496 &  1.067 & $-18.5 \pm 1.0$  & $-13.9 \pm 1.2$  & $2.77 \pm 0.71$  &     $+0.49$ & $36 \pm 15$  & $100-300$ \\
153223.20$-$083200.9 & 9.026  &         2.727 &  4.763 & $+38.5 \pm 1.3$ & $+17.69 \pm 0.40$ & $4.36 \pm 0.25$  &     $-0.63$ & $< 11$       & $100-500$ \\ 
155444.93$-$075204.6 & 11.384 &         3.151 &  5.611 & $-1.1 \pm 2.6$   & $-33.5 \pm 1.0$  & $4.73 \pm 0.81$  &     $-0.82$ & $135 \pm 22$ & $30-200$  \\
165025.84$+$272817.2 & 11.295 &         1.935 &  1.867 & $-41.0 \pm 1.1$  & $-23.9 \pm 0.7$  & $5.96 \pm 0.86$  &     $+0.54$ & $135 \pm 20$ & $30-200$  \\
172020.72$+$575837.3 & 8.951  &         1.510 &  0.256 & $-17.7 \pm 1.7$ & $+9.1 \pm 1.6^{\rm d}$ & $10.09 \pm 0.23$ &     $+1.38$ & $< 7$        & $> 1000$  \\
174311.06$+$334948.9 & 11.585 &         2.604 &  1.103 & $-27.1 \pm 1.0$  & $-13.0 \pm 2.6$  & $9.34-12.34$     &     $-0.23$ & $120 \pm 24$ & $30-200$  \\
180816.02$+$294128.1 & 8.083  &         2.477 &  1.842 & $-5.8 \pm 2.2$   & $-40.2 \pm 1.2$  & $29.09$          &     $+0.59$ & $152 \pm 22$ & $30-200$  \\
181938.10$+$364059.2 & 11.549 &         2.897 &  0.962 & $-22.0 \pm 0.4$  & $-25.0 \pm 0.8$  & $0.75 \pm 0.28$  &     $-0.91$ & $103 \pm 20$ & $100-300$ \\
184633.09$+$485444.8 & 12.127 &         3.131 &  5.100 & $-98.8 \pm 0.9$  & $-18.9 \pm 2.6$  & $10.72-14.96$    &     $-2.93$ & $165 \pm 30$ & $30-200$  \\
210124.59$+$054212.8 & 11.667 &         2.278 &  0.988 & $+22.7 \pm 0.5$  & $-5.0 \pm 0.8$   & $7.12-9.13$      &     $+0.06$ & $160 \pm 61$ & $30-200$ \\
213100.42$+$232008.6 & 9.241  &         2.861 &  0.423 & $-14.1 \pm 0.7$  & $-1.1 \pm 1.2$   & $41.31 \pm 0.22$ &     $-0.75$ & $109 \pm 30$ & $30-200$ \\
220041.59$+$271513.5 & 11.366 &         3.642 &  0.524 & $-5.32 \pm 1.0$  & $-0.3 \pm 0.5^{\rm e}$ & $23.64-35.24$    &     $-2.14$ & $< 16$       & $30-200$ \\
222803.99$+$183606.5 & 9.925  &         1.462 &  0.323 & $+42.4 \pm 1.5$  & $-35.5 \pm 0.9$  & $5.11 \pm 0.35$  &     $+1.35$ & $< 5$        & $> 1000$ \\
235952.73$+$294947.3 & 13.096 &         1.897 &  0.283 & $-17.5 \pm 4.9$  & $+10.5 \pm 3.3$  & $3.32$           &     $+0.73$ & $< 11$       & $> 1000$ \\
\hline
\end{tabular}
\end{center}
\label{T_TLSPBs}
\end{table*}
}

\bibliography{BJW2017}

\end{document}